\documentclass[manuscript]{aastex701}

\begin{document}

\title{Turbulent Diffusion of Magnetic Field Lines in the Heliosphere}

\author[orcid=/0009-0000-6881-4420,sname='Joubert']{J.~V.~A. Joubert}
\affiliation{Centre for Space Research, North-West University, Potchefstroom, South Africa}
\email[show]{jvajoubert@gmail.com} 

\author[orcid=0000-0002-0205-0808,sname='Strauss']{R.~D. Strauss}
\affiliation{Centre for Space Research, North-West University, Potchefstroom, South Africa}
\affiliation{National Institute for Theoretical and Computational Physics (NITheCS), Potchefstroom, South Africa}
\email[show]{dutoit.strauss@nwu.ac.za} 

\author[orcid=/0000-0000-0000-0000,sname='Light']{J. Light}
\affiliation{Centre for Space Research, North-West University, Potchefstroom, South Africa}
\email{juandrelight@gmail.com} 

\author[0000-0003-3659-7956]{N.E. Engelbrecht}
\affiliation{Centre for Space Research, North-West University, Potchefstroom, South Africa}
\affiliation{National Institute for Theoretical and Computational Physics (NITheCS), Potchefstroom, South Africa}
\email{n.eugene.engelbrecht@gmail.com}

\author[0000-0003-0142-8669]{N.H. Bian}
\affiliation{State Key Laboratory of Lunar and Planetary Sciences and CNSA Macau Center for Space Exploration and Science, 
Macau University of Science and Technology, Macau, China}
\email{nbian@hotmail.com}

\begin{abstract}

Due to solar wind turbulence, Parker spirals are stochastic. The dispersion of magnetic field lines is described by a convection-diffusion equation for the field line density distribution which is a function of the two heliographic angles in addition to the radial distance.  Taking  into account the radial evolution of the turbulence, the three-dimensional convection-diffusion equation is transformed into a set {of} stochastic differential equations which is solved numerically using both a forward and backward formulation. By tracing a large number of stochastic Parker spirals, the field line density distribution is constructed at any point in the heliosphere. It is shown that the angular part of the distribution function can be well-fitted by a two-dimensional Gaussian with standard deviation close to $ 25^{\circ}$ at 1 AU. The simulations also confirm that the magnetic field lines are underwound, on average, for strong enough turbulence intensity. Applying the backward approach, magnetic field lines are traced from an observer at 1 AU back to the Sun, quantifying the probability of magnetic connection when interplanetary turbulence is accounted for. It is shown that the angular uncertainty of $\sim 25^{\circ}$ is sharply reduced to $\sim 4^{\circ}$  when the field lines are traced back to the solar wind source surface from 0.25 AU.\\

\end{abstract}

\keywords{\uat{Heliosphere}{711} --- \uat{Interplanetary magnetic fields}{824} --- \uat{Interplanetary turbulence}{830}}


\section{Introduction} 
The solar wind magnetic field is a key component of solar-terrestrial connections. Charged particles are guided by magnetic fields and thus the interplanetary magnetic field geometry is essential to the forecasting of solar energetic particle (SEP) events. In the original work of \citet{parker_dynamics_1958},
the solar wind is accelerated at the Sun and emanates from a spherical source surface with a constant radial velocity. The angular separation between two nearby fluid elements emanating from the source surface is therefore a constant. Taking the magnetic field to be frozen-in the wind expanding from the rotating Sun
 yields the basic spiral geometry of the interplanetary magnetic field (HMF) lines. However, the paths represented by the nominal Parker spirals may not accurately reflect the magnetic connection between the spacecraft and the location where the accelerated particles are released. A fundamental reason is the presence of solar wind turbulence which makes the nature of the magnetic connection intrinsically statistical: field lines rooted at different footpoints can magnetically connect to the same location at 1 AU, albeit with different probabilities. More generally, the diffusion of magnetic field lines in the heliosphere can significantly impact the transport of energetic charged particles such as cosmic rays and SEPs \cite[see, e.g.,][]{strauss_perpendicular_2018,laitinen_effect_2018,laitinen_sun_2019}.\\

In this study, we build upon the work of \citet{bian_heliospheric_2024} where the dispersion of stochastic Parker spirals in the solar wind was investigated by means of numerical simulations of the convection-diffusion equation in the ecliptic plane. The investigation is here extended to {three-dimensions}, accounting for the effects of both
longitudinal and latitudinal diffusion. We derive a set of stochastic differential equations (SDEs) equivalent to the original convection-diffusion equation, which is solved numerically. The model assumes isotropic diffusion in the local coordinate system  perpendicular to the mean magnetic field \cite[see, e.g.,][]{matthaeus_spatial_1995}. The form of the angular diffusivity, an increasing function of radial distance, is constrained from observed turbulence quantities \cite[see][]{EngelbrechtEA22}. The SDEs are solved in either a `forward' or `backward' fashion. In the backward approach, individual magnetic field lines are traced from an observer, located e.g. at 1 AU, back to the solar source surface. This allows us to effectively back-map the magnetic field lines from the spacecraft to the Sun. {It is perhaps important to note that the fieldline random walk (FLRW) model employed here is based on a Markov process with independent increments and no memory effects. While this is a useful description, solar wind turbulence does not consist of purely random fluctuations, but has an underlying correlation that can give rise to coherent magnetic structures \citep[e.g.][]{Hussain_1986,Ruffolo_etal_2003,Chian_etal_2022}; a property not described by the FLRW simulations presented here. }\\

The manuscript is organized as follows. In Sec.~\ref{Sec:FLRW_model}, we present the FLRW model along with the form of the diffusion coefficient in relation to the turbulence inputs. Sec.~\ref{Sec:results} gives the results obtained from the FLRW model and discusses the methods used to analyse the obtained results. In Sec.~\ref{Sec:discussion} the implication of the results, as well as the analyses thereof, are discussed. Finally Appendix \ref{Sec:num_aspects} gives a more in-depth discussion of the mathematical and numerical aspects behind the FLRW model used in this study.\\
\section{The field line random walk model}
\label{Sec:FLRW_model}

The FLRW model is based on the focused transport equation \citep{skilling_cosmic_1971,Zhang2006} governing the time evolution of the guiding center density distribution function $f(\vec{r},\mu,t)$ of energetic particles,
\begin{eqnarray}
	\frac{\partial f}{\partial t} + {\nabla} \cdot \left(\mu v\hat{e}_bf\right) &+& \frac{\partial}{\partial\mu}\left(\frac{1-\mu^2}{2L}vf\right)\label{eq:transport equation}  \\  \nonumber &=& 
    \frac{\partial}{\partial \mu}\left(D_{\mu\mu}\frac{\partial f}{\partial \mu}\right)  + {\nabla}\cdot\left(\mathbf{D}_{p}\cdot{\nabla} f\right),
\end{eqnarray}
where $\mu$ is the pitch-angle cosine, $L$ is the magnetic focusing length, $\hat{e}_b$ is the unit vector along the guiding magnetic field and $D_{\mu\mu}$ is the pitch-angle diffusion coefficient. Magnetic focusing, a consequence of the conservation of the first adiabatic invariant, and pitch angle scattering are finite Larmor radius effects. For $\mu=1$, the spatial diffusion tensor $\mathbf{D}_{p}$ is related to the magnetic field line diffusion tensor $\mathbf{D}$ by $\mathbf{D}_{p} = v  \mathbf{D}$, where $v$ is the particle speed \citep{jokipii_cosmic-ray_1966}. Therefore, taking the zero gyroradius limit $\mu=1$ and substituting $s=vt$ in Equation (\ref{eq:transport equation}) yields the convection diffusion equation,
\begin{equation}
	\frac{\partial f_m}{\partial s}+{\nabla}\cdot\left(\hat{e}_bf_m\right)={\nabla}\cdot\left(\mathbf{D}\cdot{\nabla} f_m\right)\label{eq:convection diff}
\end{equation}
 where $f_m(\vec{r},s)$ represent the magnetic field line distribution function. In terms of a local orthonormal field-aligned coordinate system with the first axis parallel to the guiding magnetic field, the diffusion tensor is diagonal and given by
\begin{equation}
\textbf{D}=\left[\begin{array}{c c c}
     0 & 0 & 0  \\
     0 & D_{\perp} & 0 \\
     0 & 0 & D_{\perp}
\end{array}\right],
\end{equation}
for a FLRW process in which only diffusion perpendicular to the guiding magnetic field is considered. {Note that we assume perpendicular diffusion which is isotropic with respect to the guiding magnetic field, and hence only specify a single perpendicular diffusion coefficient $D_{\perp}$.} The guiding magnetic field is taken to be the \citet{parker_dynamics_1958} field,
\begin{equation}
	\vec{B}(r,\theta) = B_0\left(\frac{r_0}{r}\right)^2(\hat{e}_r - \tan\psi\hat{e}_{\phi})\label{eq:parker},
\end{equation}
where $r$ is the radial distance from the Sun, $\phi$ is the heliolongitude, $\theta$ is the heliolatitude and $B_0$ is the field magnitude at the reference radius $r_{0}$. The winding angle $\psi$, the angle between $\hat{e}_b$ and $\hat{e}_r$, satisfies
\begin{eqnarray}
    \tan\psi = \frac{\Omega (r-r_0)\sin\theta}{v_{sw}},
\end{eqnarray}
where $\Omega= 14.1844^{\circ}$/day is the equatorial rotation rate of the Sun and $v_{sw}=400$ km/s is the solar  wind speed. The solar wind magnetic field is assumed to emanate from a source surface located at the radial distance $r_0=0.04$ AU.

Equation (\ref{eq:convection diff}) can be solved using a set of SDEs for individual (notional) magnetic field lines traced either forward from the Sun into the heliosphere or backward from the spacecraft to the Sun. A detailed derivation of these SDEs is provided in the Appendix \ref{Sec:num_aspects}. In the forward approach, the convection-diffusion equation is first transformed into a Fokker-Planck (Kolmogorov forward) equation governing the evolution of the probability density function (PDF) $g_{m}(r,\theta, \phi,s)=f_m(\vec{r},s)r^2\sin\theta$, 
\begin{equation}
	\frac{\partial g_m}{\partial s} = -\sum_{i=1}^{3}\frac{\partial}{\partial x_i}\left[a_ig_m\right]+\frac{1}{2}\sum_{i=1}^{3}\sum_{j=1}^{3}\frac{\partial^2}{\partial x_i\partial x_j}\left[C_{ij}g_m\right],
\end{equation}
 where $a_i$ is the drift vector{, containing all convective terms,} and $C_{ij}$ is the diffusion tensor. {In the SDE literature, it has become custom to refer to the convective terms as ``drift'' terms, but this should not be conflated with the physical process of charged particle drifts, e.g. gradient and curvature drifts.} Introducing a positive definite matrix $b_{ij}$, related to the diffusion tensor $C_{ij}$ by $b_{ij} \cdot b_{ij}^T = C_{ij}$, the Fokker-Planck equation can then be expressed in terms of an equivalent set of SDE's,  
\begin{equation}\label{sps}
    \frac{dx_i(s)}{ds} = a_i(x_{i}) + \sum_{j=1}^{3}b_{ij}(x_{i}) \, \xi_i(s),
\end{equation}
where  $\xi_{i}(s)=dW_{i}(s)/ds$ is the Gaussian white noise. The expressions for the drift vector and the diffusion tensor in the spherical coordinate system $(r,\theta,\phi)$ are derived in the Appendix \ref{Sec:num_aspects}. The matrix $b_{ij}$
is related to the magnetic field line diffusion coefficient $D_{\perp}(r)$, a function of the radial distance $r>r_{0}$, by
\begin{eqnarray}
\textbf{b} = \sqrt{2 D_{\perp}(r)}\left[\begin{array}{ccc}
		0 & \qquad0 & \qquad \sin\psi\\
		0 & \qquad \frac{1}{r}& \qquad 0\\
		0 & \qquad 0 & \qquad \frac{\cos\psi}{r\sin\theta}
	\end{array}\right],
\end{eqnarray}
The form of $D(r)$ adopted in this study will be discussed in the next section. The solutions [$r(s),\theta(s), \phi(s)$] of Eq. (\ref{sps}) are stochastic Parker spirals.
Setting $D_{\perp}=0$ in Eq. (\ref{sps}), they become the solutions of the ordinary differential equations,
\begin{eqnarray}
    \frac{dr}{ds} &=& \cos \psi,  \nonumber \\
\frac{d\theta}{ds} &=& 0,\\
    \frac{d\phi}{ds} &=& -\frac{\Omega}{v_{sw}} \cos \psi, \nonumber 
\end{eqnarray}
recovering the nominal Parker spirals  $[\phi(r) = - \Omega (r-r_{0})/ v_{sw} + \phi_0, \theta(s)=\theta_{0}]$. Here $(\theta_{0},\phi_{0})$ is the position of the magnetic footpoint on the solar wind source surface located at the radial distance $r_{0}$. Using the Kolmogorov backward equation, another set of SDE's can be obtained which traces the notional magnetic field lines from the observer's position back to the solar wind source surface. 

\subsection{Field line diffusion coefficient}

Assuming a composite model of transverse turbulence \cite[see][]{matthaeus_evidence_1990}, the diffusion coefficient is given by \cite[e.g.][]{matthaeus_spatial_1995,ruffolo_separation_2004,chuychai_trapping_2007,strauss_perpendicular_2018,shalchi_field_2019,EngelbrechtEA22},
\begin{equation}
    D_{\perp} = \lambda_u\sqrt{\frac{\delta B^{2}_{2D}	}{2B^{2}_0}} \label{eq:diff coeff},
\end{equation} 
where $\delta B^{2}_{2D}$ is the magnetic variance and $\lambda_u$ is the ultrascale, the typical size of magnetic islands in the two-dimensional turbulence model \citep{matthaeus_spectral_2007}. The spatial dependences of $D_{\perp}$ arise from those of the turbulence quantities (and $B_0$) that it is a function of. The power spectrum is taken in the form given by \citet{engelbrecht_sensitivity_2015},
\begin{eqnarray}
S_{2D}(k_{\perp}) & = & \frac{C_{0}\lambda_{2D}\delta B^{2}_{2D}}{2\pi k_{\perp}} \left\{ \begin{array}{ll}
(\frac{\lambda_{2D}}{\lambda_o})^{-1}(\lambda_ok_{\perp})^{q}  \textrm{, }|k_{\perp}|<{\lambda_o^{-1}}; \\
(\lambda_{2D}k_{\perp})^{-1}  \textrm{, }\lambda_o^{-1}\le|k_{\perp}|<\lambda_{2D}^{-1};\\
(\lambda_{2D}k_{\perp})^{-\nu}  \textrm{, }|k_{\perp}| \ge \lambda_{2D}^{-1},
\end{array}\right. \nonumber
\label{eq:22D}
\end{eqnarray}
It consists of three ranges: the inertial range at the highest wavenumbers with a spectral index $\nu$, assumed here to be equal to the Kolmogorov value \cite[see, e.g.,][]{LotzEA23} and commencing at lengthscale $\lambda_{2D}$; an energy-containing range at intermediate wavenumber with an observed $k^{-1}$ wavenumber dependence \cite[see, e.g.,][]{MattEA07b,WangEA24,PradataEA25} commencing at $\lambda_o$; and an inner range at the smallest wavenumbers, with a spectral index of $q=3$ chosen so as to ensure a finite ultrascale while ensuring that the solenoidal condition is satisfied \cite[see][]{matthaeus_spectral_2007}. The constant $C_0$ is obtained from the normalization $\int S_{2D}(k_{\perp})dk_{\perp}=\delta B^{2}_{2D}$, yielding
\begin{equation}
C_{0}=\left[\left(\log \left(\frac{\lambda_o}{\lambda_{2D}}\right)+\frac{1}{1+q}+\frac{1}{\nu-1}\right)\right]^{-1}.
\label{eq:Cmin1}
\end{equation}
The ultrascale entering Equation (\ref{eq:diff coeff}) is thus given by
\begin{equation}
    \lambda_u = \sqrt{C_0 \left[\lambda_{o}^{2}\left(\frac{q+1}{2q-2}\right)+\lambda_{2D}^{2}\left(\frac{1-\nu}{2\nu+2}\right)\right]}.
\end{equation}
Here, we assume that the 2D magnetic variance represents $80$\%\footnote{This fraction is here assumed to remain constant as function of radial distance in the very inner heliosphere \cite[see observations reported by][]{Chhiber22}.} of the total variance \cite[although this has been observed to vary, see, e.g.,][]{BieberEA94,BieberEA96,oughton_anisotropy_2015}, which is modeled using a power-law radial dependence motivated by spacecraft observations \cite[see][]{zank_evolution_1996,BurgerMcKee23} as well as the results of turbulence transport models \cite[see, e.g.,][]{breech_turbulence_2008,OughtonEA11,EngelbrechtStrauss18,AdhikariEA21} such that
\begin{equation}
    \delta B^2_{2D} = 12.5\left(\frac{r}{r_e}\right)^{-2.4} (\textrm{in }\mathrm{nT}^2) ,
\end{equation}
normalized to a solar minimum observational value at $r_e = 1$ AU from \citet{smith_dependence_2006}. The lengthscale at which the inertial range commences is modeled following the behaviour of the 2D correlation scale. As such, a simple power law radial dependence motivated by spacecraft observations both within and beyond $1$~AU \cite[see, e.g.,][]{smith_heating_2001,cuesta_isotropization_2022} is chosen,
\begin{equation}
     \lambda_{2D} = 0.0074\left(\frac{r_e}{r}\right)^{1/2} (\textrm{in }\mathrm{AU}), 
 \end{equation}
which is normalized to a value at Earth reported by \citet{weygand_correlation_2011}. Note that no solar cycle dependence for these quantities is modeled here, and, as such, the above expressions are normalized to solar minimum values at $r_e$ \cite[see][]{ZhaoEA18,BurgerEA22}. Furthermore, both $\delta B^2_{2D}$ and $\lambda_{2D}$ could reasonably be expected to display a latitudinal dependence, from observations and turbulence transport modelling further out in the heliosphere \cite[see, e.g.,][]{ForsythEA96,Bav00a,Bav00b,breech_turbulence_2008,EngelbrechtBurger13}, but it is unclear what this dependence would be at the radial distances relevant to this study. As a first approach, a purely radial dependence based on extant observations is assumed. Modeling the lengthscale at which the energy-containing range begins can be problematic as no direct observations for this quantity exist. As such, we employ a scaling based on particle transport model \citep{Engelbrecht19outer},
\begin{equation}
\lambda_o = 12.5 \lambda_{2D}
\end{equation}
which is close to the frequency range over which the energy containing range extends from observations \citep{WangEA25}.

\begin{figure*}[!t]
\includegraphics[width=0.49\textwidth]{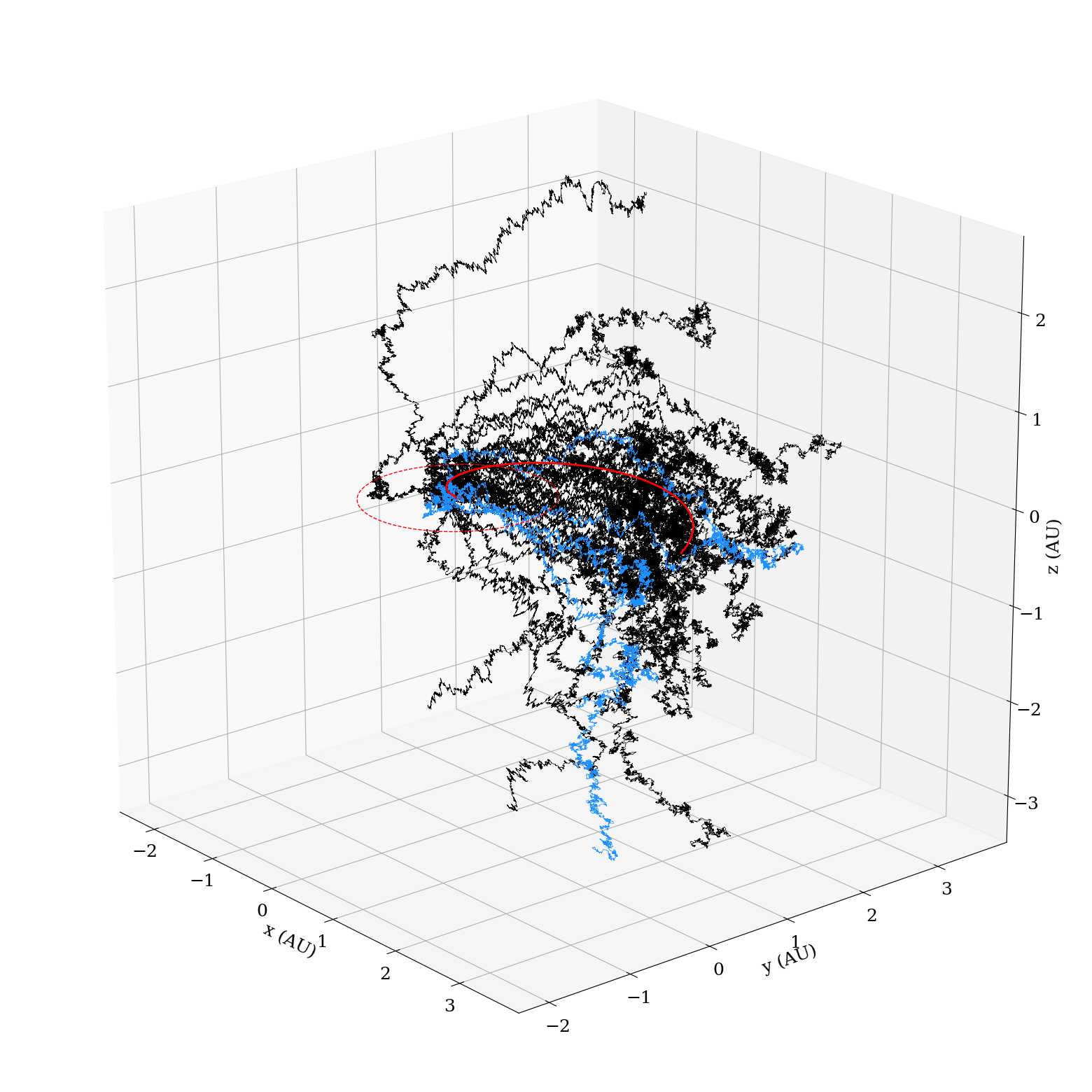}
\includegraphics[width=0.49\textwidth]{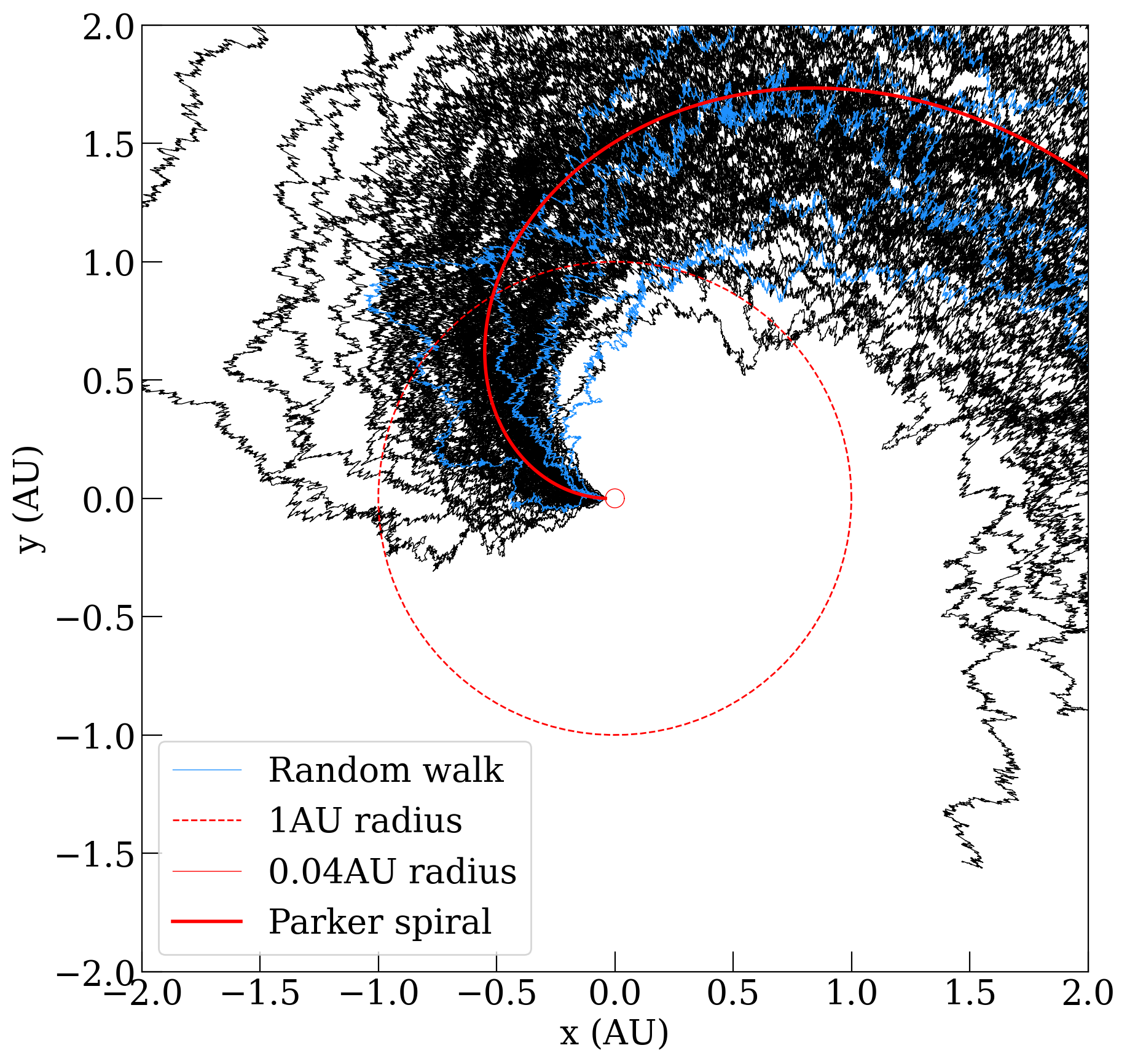}
\caption{Left: 100 stochastic Parker spirals (solid black and blue curves) along with the nominal Parker spiral (thick solid red curve) emanating from the same magnetic footpoint on the solar wind source surface. Right: Their projection on the ecliptic plane. The dashed red circle has a radius of 1 AU. The thin solid red circle has a radius $r_{0}=0.04 AU$ corresponding to the source surface radius from which all the magnetic field lines emanate. \label{fig:random walk}}
\end{figure*}

\section{Simulation results}\label{Sec:results}
\subsection{Forward simulations}

Let us first present the results of the forward simulations where the  field lines are traced from the Sun into the heliosphere. Fig.~\ref{fig:random walk} displays 100 stochastic Parker spirals solutions of Eq.~(\ref{sps}). They all share a common magnetic footpoint on the source surface of radius $r_{0}=0.04$ AU. Few magnetic field lines are plotted in blue in order to better depict the stochastic nature of each solution curve. It can be clearly seen that the magnetic field lines tend to disperse, in both heliographic longitude and latitude, away from the nominal Parker spiral which is plotted in red. These stochastic Parker spirals are different realizations of the same stochastic process representative of the statistical ensemble \citep{shalchi_field_2019}. As emphasized by   \citet{bian_stochastic_2021}, stochastic Parker spirals correspond to the paths followed by scatter-free charged particles injected on the magnetic footpoint with a pitch angle cosine $\mu=1$ in this solar wind turbulence model. By analogy with the nomenclature adopted in Monte Carlo simulations of particle transport, such stochastic Parker spirals can also be referred to as ``pseudo'' or notional magnetic field lines; these curves are not divergence free and, although they share some characteristics of turbulent magnetic field lines, are not real magnetic field lines. They are used here to compute the probability density $g_{m}(r,\theta,\phi,s)$ solution of the Fokker-Planck equation and thus, to reconstruct the field line density $f_{m}(r, \theta, \phi, s)$  at any point in the heliosphere. To calculate the probability distribution of magnetic field lines, a spatial grid is used and when a SDE solution passes through a specific grid element, a corresponding weight is added to the distribution function, which is, at the end, divided by the number of solutions passing through the grid element.

\begin{figure*}[!t]
\includegraphics[width=0.49\textwidth]{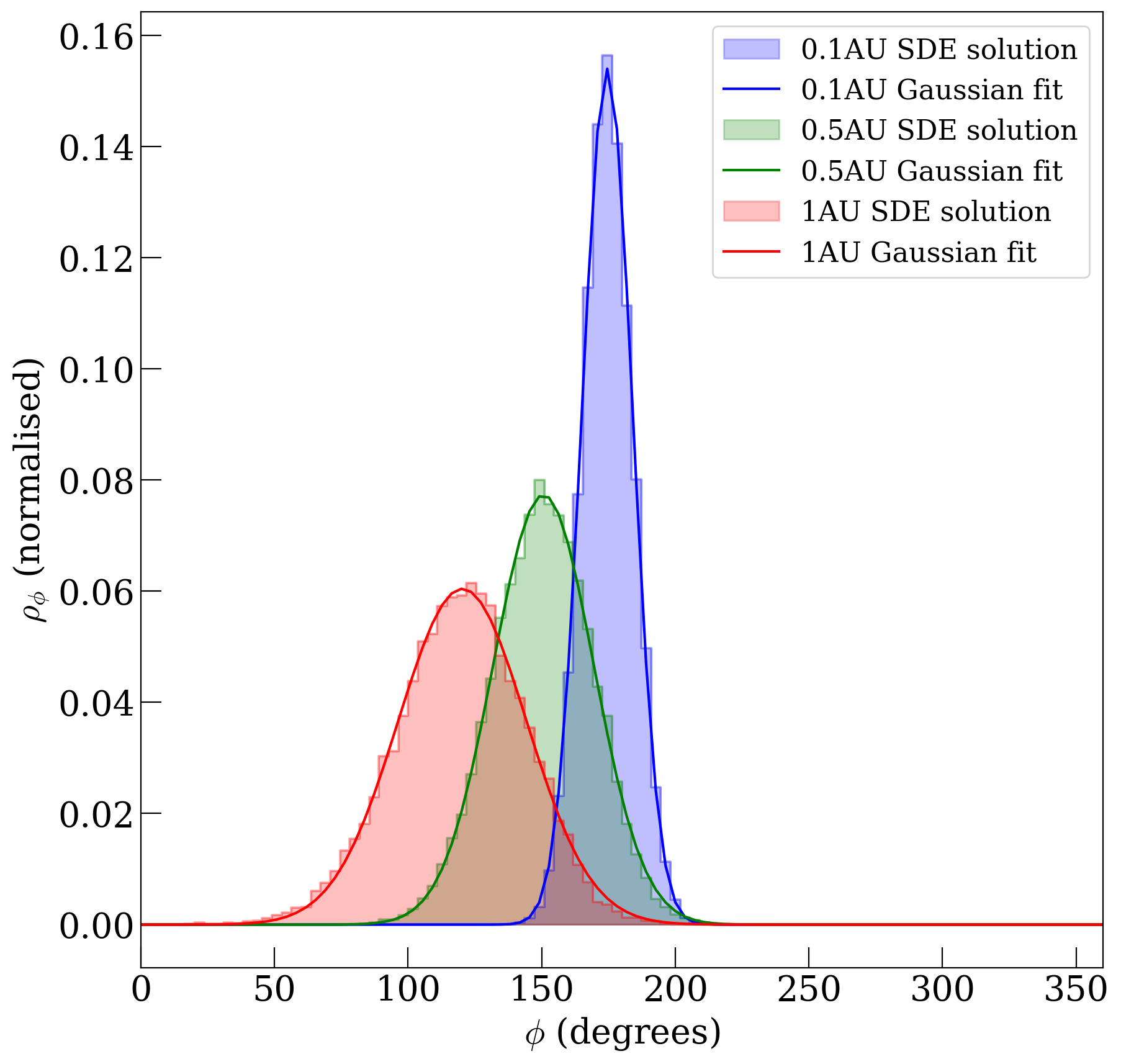}
\includegraphics[width=0.5\textwidth]{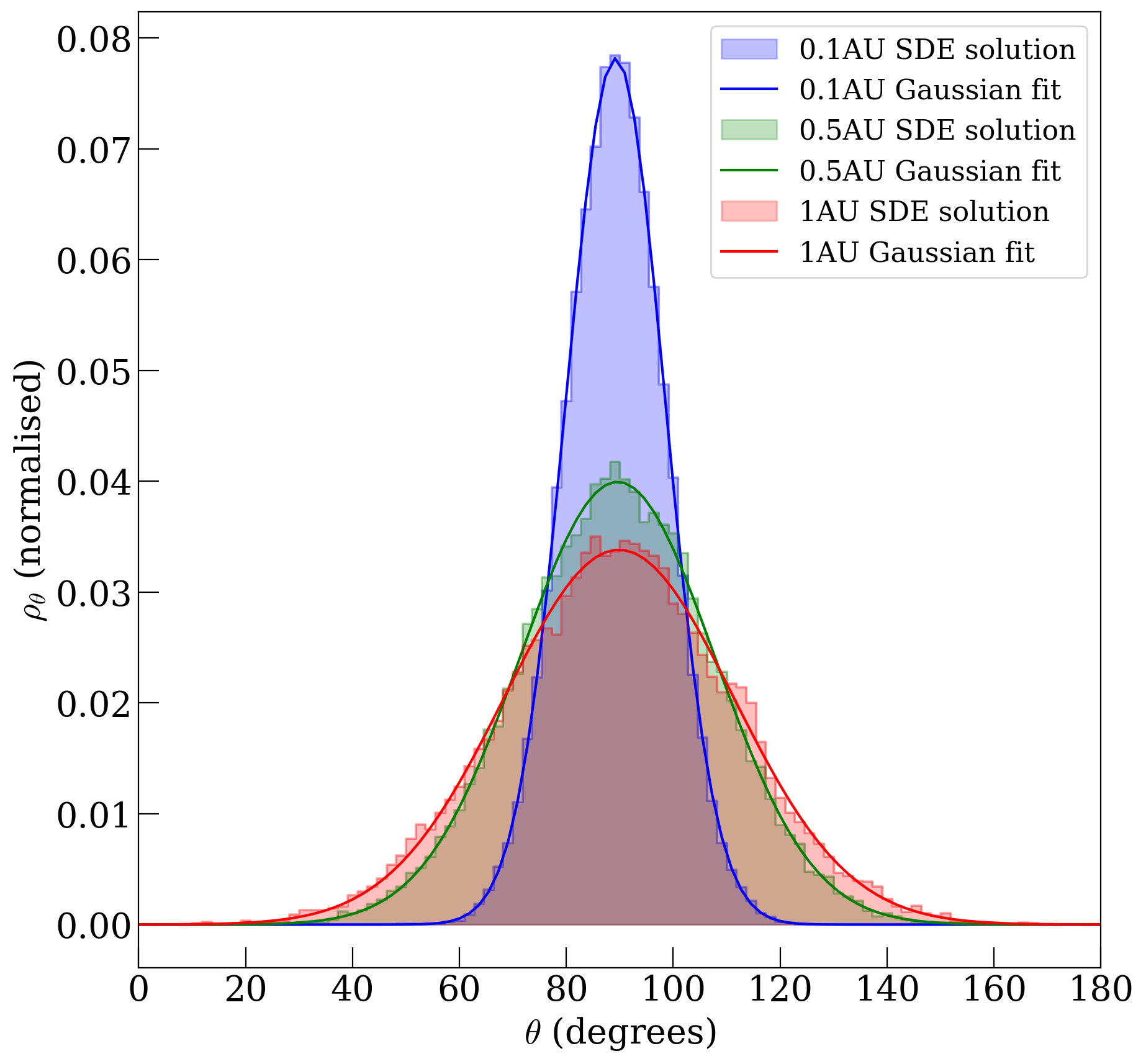}
\caption{Angular field line distribution, as a function of $\phi$ (left) and $\theta$ (right), compared to Gaussian fits at radial distances $0.1$ AU (blue), $0.5$ AU (green) and 1 AU (red) from the Sun. \label{fig:hist}}
\end{figure*}

\begin{figure}[!t]
\includegraphics[width=0.49\textwidth]{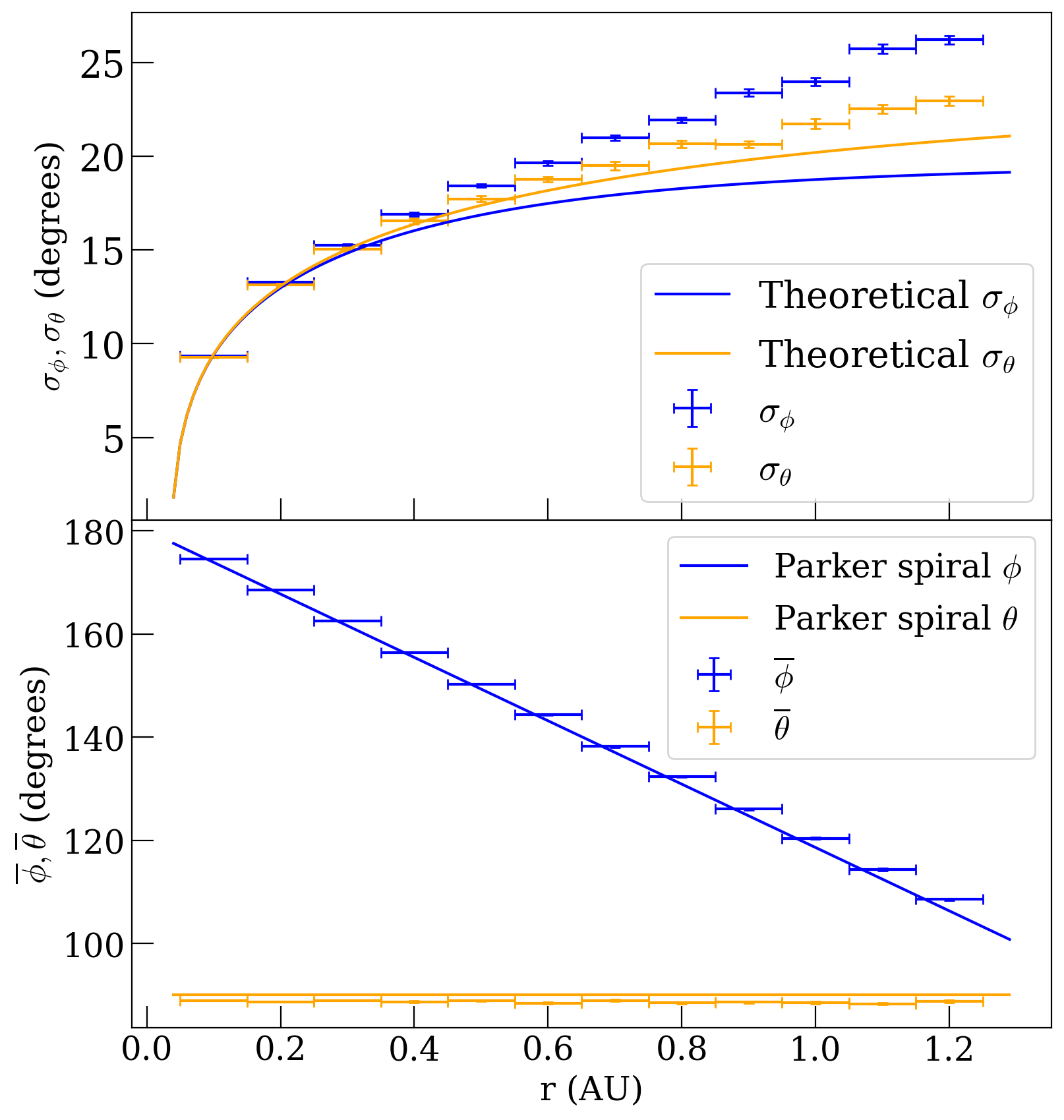}
\caption{\textit{Top}: $\sigma_{\phi}(r)$ (blue crosses) and $\sigma_{\theta}(r)$ (orange crosses) determined from the Gaussian fit
to the angular distribution. They are
compared to the model for $\sigma_{\phi}(r)$ (solid blue line) and $\sigma_{\theta}(r)$ (solid orange line) given by Eq.~(\ref{thsig}). \textit{Bottom}: $\overline{\phi}(r)$ values (blue crosses) and $\overline{\theta}(r)$ (orange crosses) compared to the Parker spiral $\phi(r)$ (solid blue line) and $\theta(r)$ (solid orange line) at interval distances of 0.1 AU from the Sun. \label{fig:stand dev}}
\end{figure}

As it appears in Fig.~\ref{fig:hist}, the angular distribution of magnetic field lines can be well-fitted by the two-dimensional Gaussian distribution,
\begin{equation}
    f_{a}(\theta,\phi) = \frac{1}{2\pi\sigma_{\theta}(r)\sigma_{\phi}(r)} e^{-\left[\frac{(\theta- \overline{\theta}(r))^2}{2\sigma_{\theta}^2(r)}+\frac{(\phi-\overline {\phi}(r))^2}{2\sigma_{\phi}^2(r)}\right]},
\end{equation}
 as expected in the small angle approximation of an angular diffusion  process. Therefore, we make the ansatz that the angular distribution of magnetic field lines around its mean value obeys the diffusion equation
\begin{equation}\label{conts2}
\frac{\partial f_{a}}{\partial r} =\frac{\partial }{\partial \Delta \theta}D^{\theta\theta}_{a}(r)\frac{\partial f_{a}}{\partial \Delta \theta}+\frac{\partial }{\partial \Delta \phi}D^{\phi\phi}_{a}(r)\frac{\partial f_{a}}{\partial \Delta \phi},
\end{equation}
where the coefficients $D^{\phi\phi}_{a}(r)$ and $D^{\theta\theta}_{a}(r)$ represent the longitudinal and latitudinal diffusivities, respectively. The angular diffusivities are related to the components $D_{\phi \phi}(r)$ and $D_{\theta \theta}(r)$ of the diffusion tensor by 
\begin{equation}
D^{\phi\phi}_{a}(r)=\frac{D_{\phi\phi}(r)}{r^{2}},
\end{equation}
with a similar equation for $D^{\theta\theta}_{a}(r)$. It follows from these model assumptions that the standard deviations of the Gaussian distribution are predicted to be given by 
 \begin{eqnarray}\label{thsig}
    \sigma_{\phi}(r) = \sqrt{2\int_{r_0}^rD^{\phi\phi}_{a}(r^{\prime})dr^{\prime}},
 \end{eqnarray}
 and a similar equation for $\sigma_{\theta}(r)$. The above model for $\sigma_{\phi}(r)$ and $\sigma_{\theta}(r)$ is compared with results of the Gaussian fit to the numerical results in Fig.~\ref{fig:stand dev}. They display a similar behavior in which the standard deviation increases quickly before reaching a plateau. The most significant difference is the larger increase of $\sigma_{\phi}(r)$ obtained from the fit causing a substantial deviation from the model predictions at radial distances $\geq 0.5$ AU. However, the theoretical prediction $\sigma_{\theta}$ appears to match the fit reasonably well. The discrepancy between the prediction of $\sigma_{\phi}(r)$ and the numerical results is due to the effect of the curvature of the Parker spiral away from the radial direction not accounted for in the theoretical model.  The bottom panel in Fig.~\ref{fig:stand dev} displays the radial evolution of $\overline{\phi}(r)$ and $\overline{\theta}(r)$ obtained from the Gaussian fit to the numerical simulations. They tend to follow $\phi(r)$ and $\theta(r)$ associated with the Parker spiral. The small error margins derived from the fitting algorithm, increase with radial distance, while the error in radial distance is simply the radial increment.\\

\begin{figure*}[!t]
\begin{center}
\includegraphics[width=0.85\textwidth]{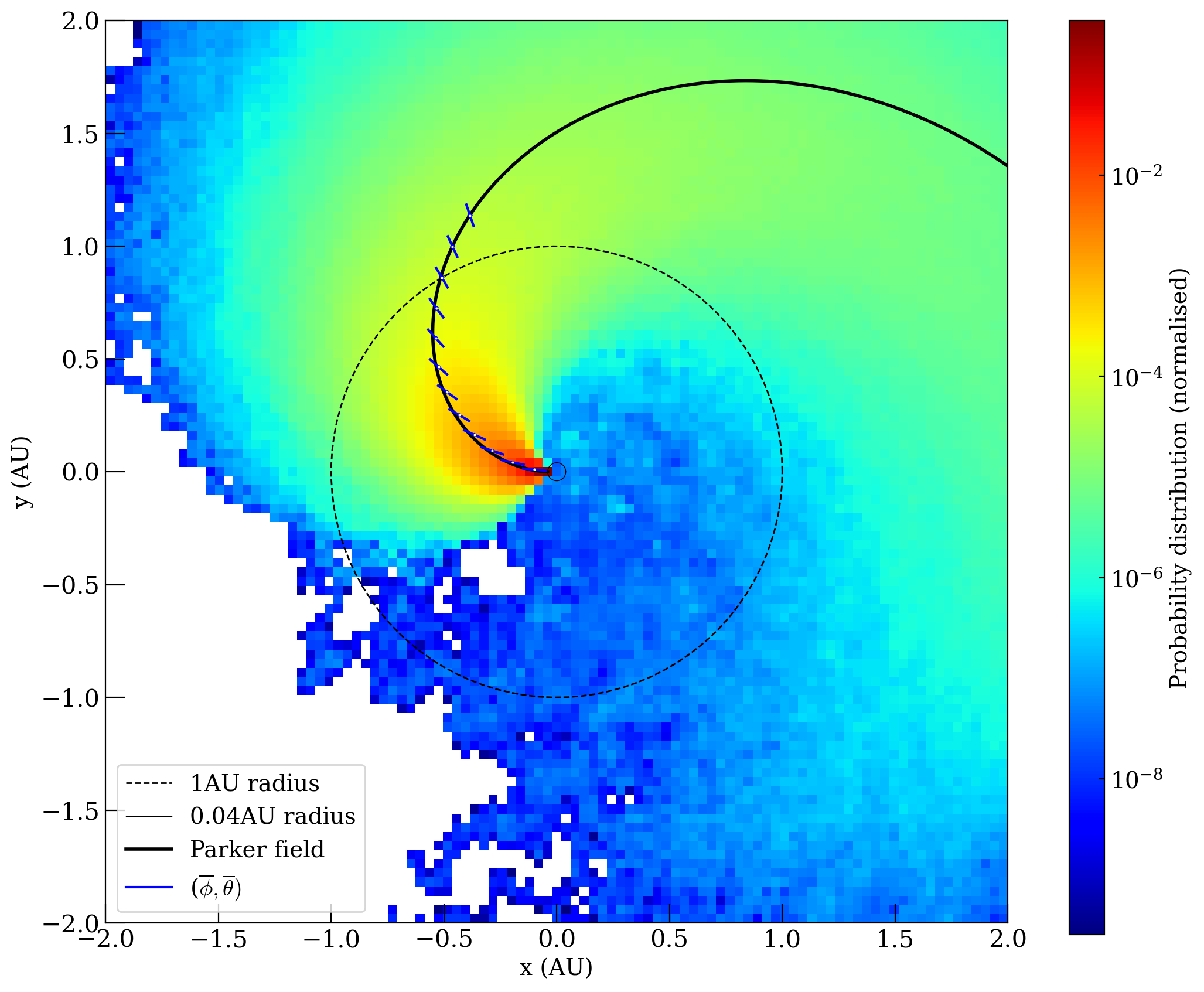}    
\end{center}
\caption{The PDF $\rho_{m}$ Density map of the 2D profile showing the normalized probability distribution calculated along the solution curves, the $(\overline{\phi},\overline{\theta})$ values (blue crosses) compared to the Parker spiral (thick solid black line), the $1$ AU radius (dashed black line) and the starting radius (thin solid black line). \label{fig:2D B-field grid}}
\end{figure*}

In Fig.~\ref{fig:2D B-field grid} the respective $x$- and $y$-coordinates of each solution curve are determined for every step along the solution curve. These coordinates are in turn used to determine the position of the solution curves on a 2D grid, which ranges from $-2$ AU to $2$ AU along both the $x$- and $y$-axis, divided into intervals of size $0.02$ AU. If a solution curve is found to be passing through any given interval, the probability distribution, as calculated by the solution curve, is added to said interval and displayed using a density map. On this logarithmic colour mapping, red represents a high probability, blue a low probability, and no colour indicates that no solution curve passed through an interval. In addition, the $(\overline{\phi},\overline{\theta})$ values determined using the Gaussian fits are illustrated in Fig.~\ref{fig:2D B-field grid} at every $0.1$ AU interval to better compare the $(\overline{\phi},\overline{\theta})$ values, and therefore the most likely path which a magnetic fieldline would follow, with the ideal Parker magnetic field. The $(\overline{\phi},\overline{\theta})$ are displayed using crosses, which represent the margin of error for the radial distance and the azimuthal angle. It should also be mentioned that white dots were added to the crosses to accentuate the centre of each cross.
\\
\begin{figure*}[!t]
\includegraphics[width=\textwidth]{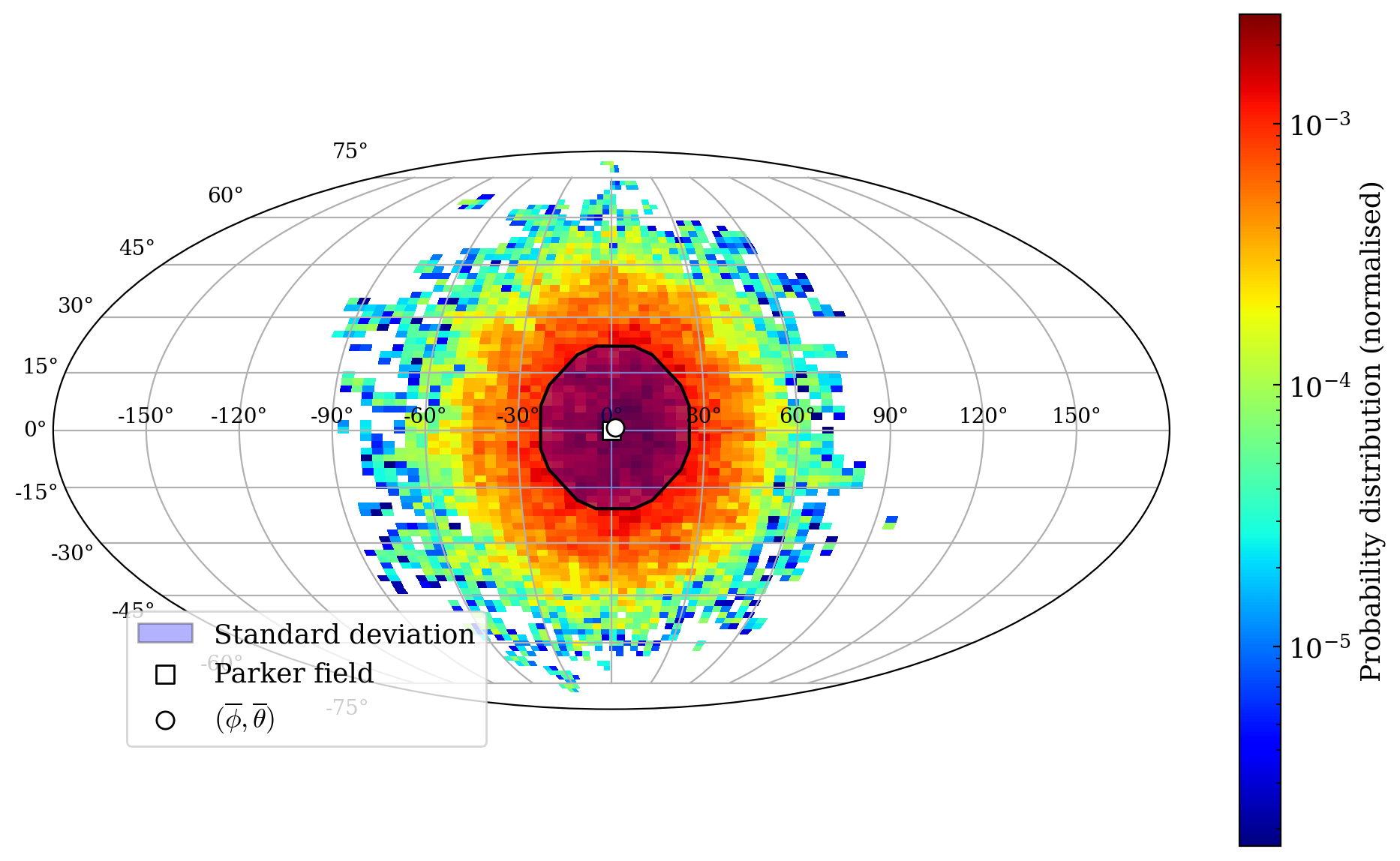}
\caption{Mollweide projection depicting the shifted probability distribution at a radial distance of 1 AU from the Sun. The $(\overline{\phi},\overline{\theta})$ value (white circle), Parker spiral (white square) and $\sigma$ (blue area) are shown as well. \label{fig:mollweide}}
\end{figure*}

Fig.~\ref{fig:mollweide} shows a Mollweide projection of the probability distribution as a function of $\theta$ and $\phi'$ at a radius of 1 AU from the Sun. Here the distribution is shifted in $\phi$ by subtracting the position of a Parker HMF at 1 AU, $\phi' = \phi - \phi_{\mathrm{Parker}}$. Similar to Fig.~\ref{fig:2D B-field grid}, the normalized probability distribution for each interval is determined and displayed using a density map, where blue once again indicates a lower intensity and red indicates a higher intensity. Along with the density map, the location of the most likely path with respect to the Parker spiral is included. Finally the standard deviation on the $(\overline{\phi},\overline{\theta})$ value at 1 AU is shown as a blue area. For magnetic fieldlines, all starting from the same exact position at the inner boundary, Fig.~\ref{fig:mollweide} gives the probability of finding a fieldline at a specific angular position at 1 AU, given the assumed level of turbulence (i.e. fieldline diffusion) in the model.
\\\\
To determine the level of meandering of the HMF due to changes in the magnitude of the diffusion coefficient, the simulation is run four more times. For these runs the point of origin for the solution curves is once again at a radial distance of $0.04$ AU, a polar angle of $\theta={\pi}/{2}$ and an azimuthal angle of $\phi=\pi$. The solar wind speed, solar rotation rate and the initial magnetic field intensity values are all identical to the reference scenario as well. The only difference between the reference scenarios and the following simulations is an {\it ad hoc} change in the magnitude of the diffusion coefficients. For simplification, the different simulations will be referred to as simulations 1 to 4. For simulation 1 the diffusion coefficient utilized in the reference scenario is assumed to be ten times smaller; for simulation 2 the diffusion coefficient is two times smaller; simulation 3 implements a diffusion coefficient two times greater than that of the reference scenario, and finally simulation 4 uses a four times greater diffusion coefficient.\\

\begin{figure}[!t]
\includegraphics[width=0.49\textwidth]{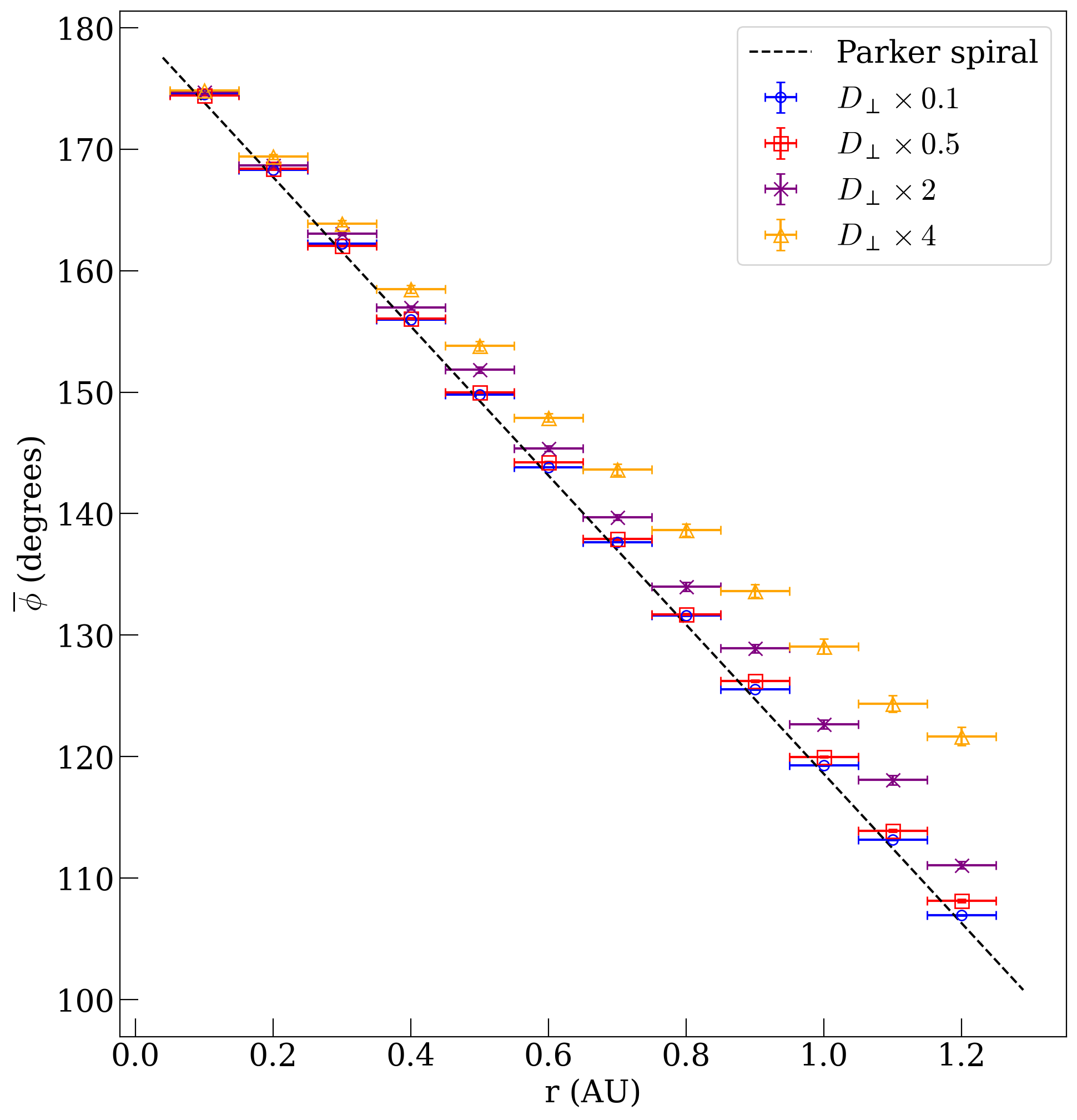}
\caption{$\overline{\phi}(r)$ values as a function of radial distance for $D_{\perp}\times4$ (orange with triangle), $D_{\perp}\times2$ (purple with cross), $D_{\perp}\times0.5$ (red with square) and $D_{\perp}\times0.1$ (blue circle). Included as well are the $\phi$ values of the ideal Parker spiral (dashed black line) as a function of radial distance. \label{fig:multi mean}}
\end{figure}

As with Fig.~\ref{fig:stand dev}, the $\overline{\phi}(r)$ values obtained from the four simulations are displayed at every $0.1$ AU interval and compared to the $\phi$ values of the Parker spiral in Fig.~\ref{fig:multi mean}. For simulations 1 and 2 the $\overline{\phi}(r)$ values remain near the Parker spiral $\phi$ values, similar to that of Fig.~\ref{fig:stand dev} and are even closer to the Parker spiral than those of the reference scenario. Simulation 3 shows the same straight line behaviour in the $\overline{\phi}(r)$ values as simulations 1 and 2, although these values deviate to a greater degree than that of simulations 1 and 2. Finally, simulation 4 shows the largest deviation in the $\overline{\phi}(r)$ values out of the four simulations. Similar to the previous simulations, simulation 4 starts with a linear decrease in the $\overline{\phi}(r)$ values, but at around 1 AU the $\overline{\phi}(r)$ values begin to decrease at a slower rate; these simulations indicate a magnetic field that is increasingly underwound relative to the Parker HMF as diffusion coefficients increase.\\

\begin{figure*}[!t]
\includegraphics[width=\textwidth]{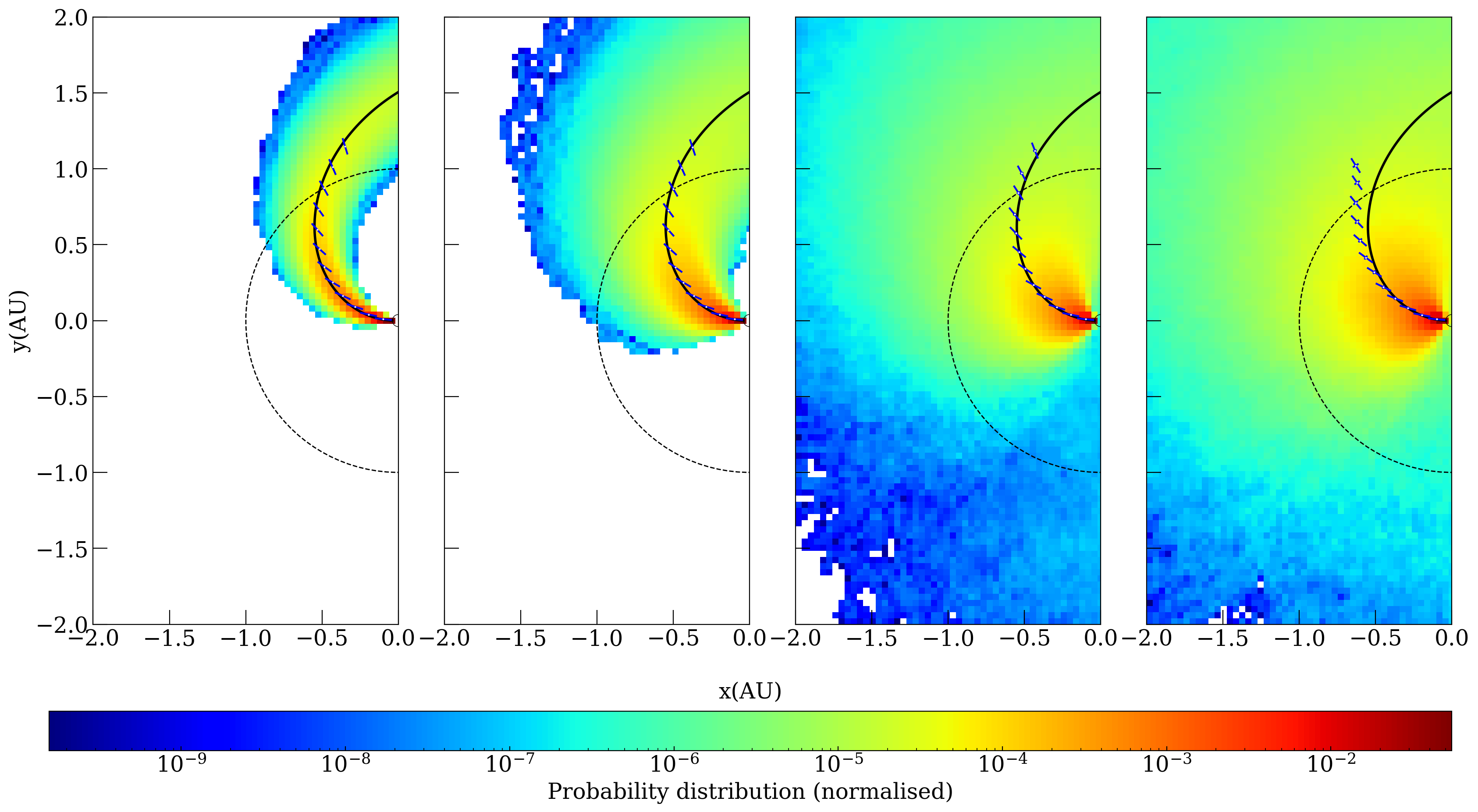}
\caption{Probability distribution, obtained from the function $g_m$, shown as intensity maps for $D_{\perp}\times0.1$ (first), $D_{\perp}\times0.5$ (second), $D_{\perp}\times2$ (third) and $D_{\perp}\times4$ (fourth panel). All of the graphs show the starting radius (solid black line), the $1\,$AU radius (black dashed line), the ideal Parker spiral (thick black line) and the most likely path (blue crosses) for each simulation. \label{fig:multi}}
\end{figure*}

Fig.~\ref{fig:multi} shows the normalized probability distribution for simulations 1 to 4 as density maps. The Parker spiral is depicted on each of the graphs as well, along with the 1 AU and point of origin radii. Finally, the most likely path for each simulation is indicated on the respective graphs as blue crosses. This value is, once again, derived from the Gaussian fits to the probability distribution where the maximum (or peak) of the distribution is assumed to give the most likely (i.e. highest probability) position of a stochastic fieldline.\\ 

\subsection{Backward simulations}

\begin{figure*}[!t]
\includegraphics[width=0.49\textwidth]{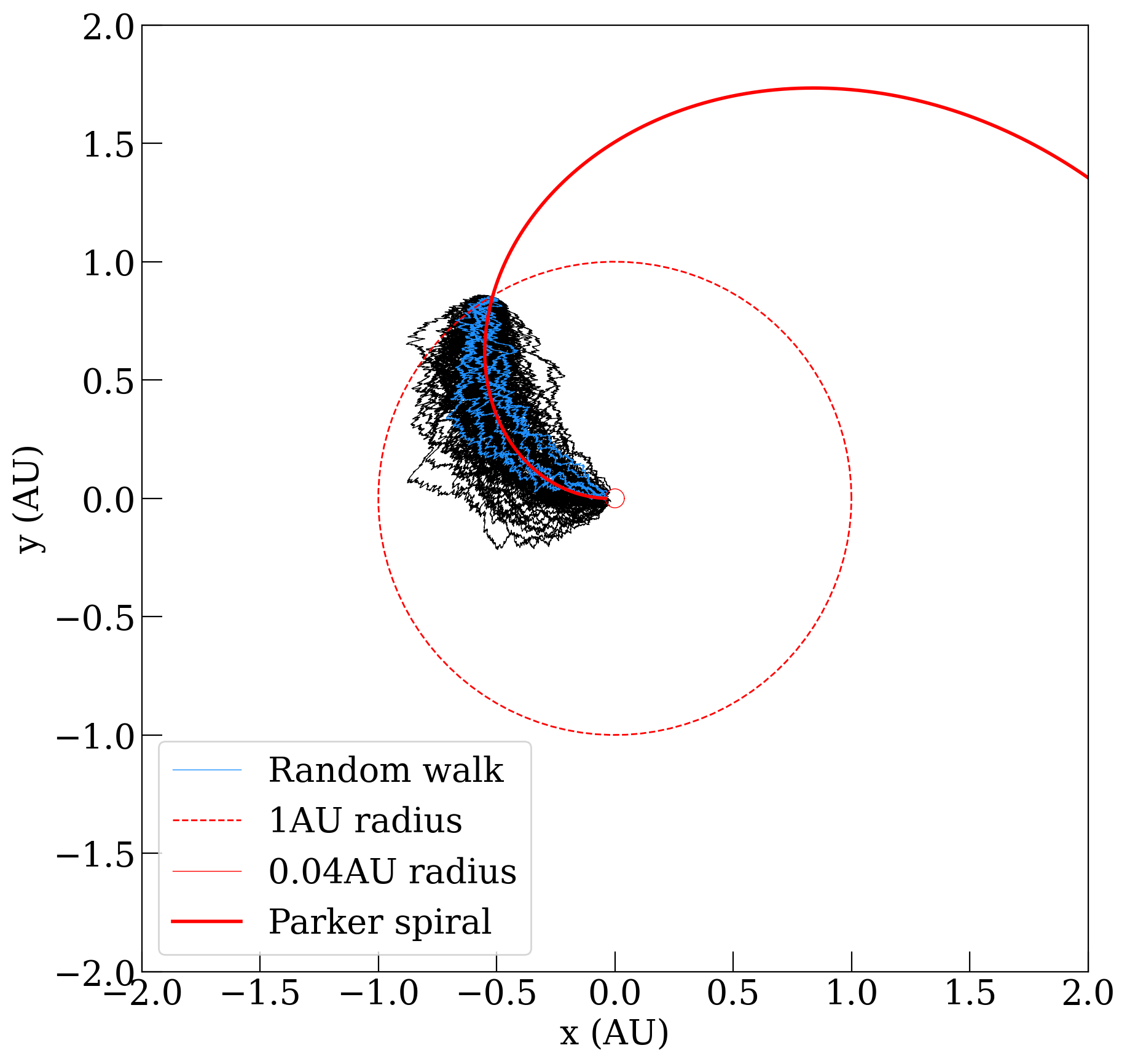}
\includegraphics[width=0.49\textwidth]{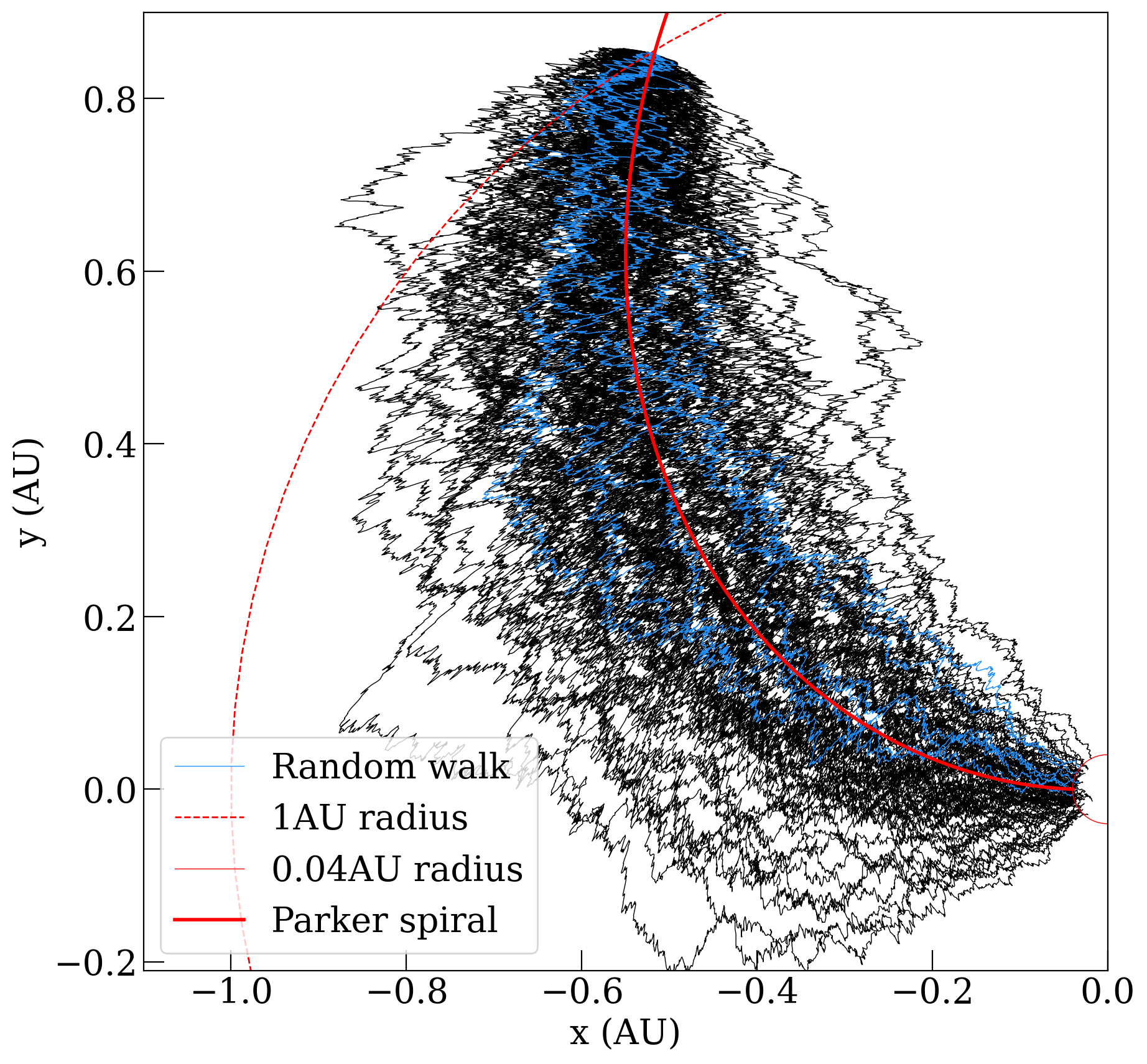}
\caption{\textit{Left}: 2D profile for 100 solution curves (solid black and blue lines), obtained using the time-backward formulation, along with the Parker spiral (thick solid red line), the 1 AU radius (dashed red line) and the end radius (thin solid red line). \textit{Right}: Zoomed in view of the 100 solution curves \label{fig:backwards}}
\end{figure*}

Fig.~\ref{fig:backwards} depicts a hundred solution curves obtained using the backwards formulation given by Eq.~\ref{eq:backwards formulation} (black lines), while the last five solution curves are depicted in blue to once more better portray the stochastic nature of each curve. Fig.~\ref{fig:backwards} displays the starting position at 1 AU, and follows the SDE solution until it reaches the inner radius of $0.04$ AU. Again the solution curves all originate from the same location and immediately start to diffuse, staying near the unperturbed Parker HMF (indicated by the thick red line). \\

\begin{figure*}[!t]
\includegraphics[width=0.5\textwidth]{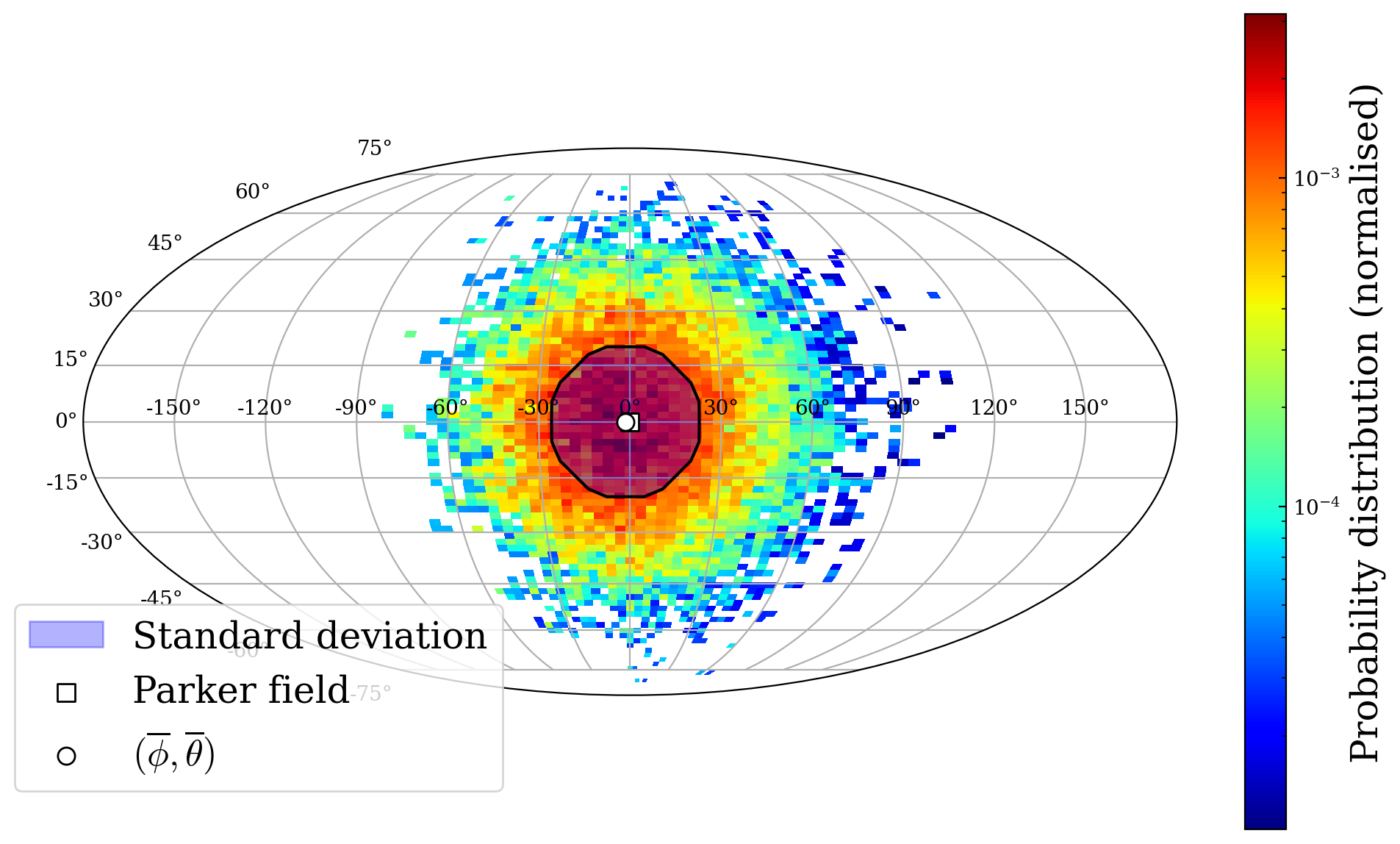}
\includegraphics[width=0.5\textwidth]{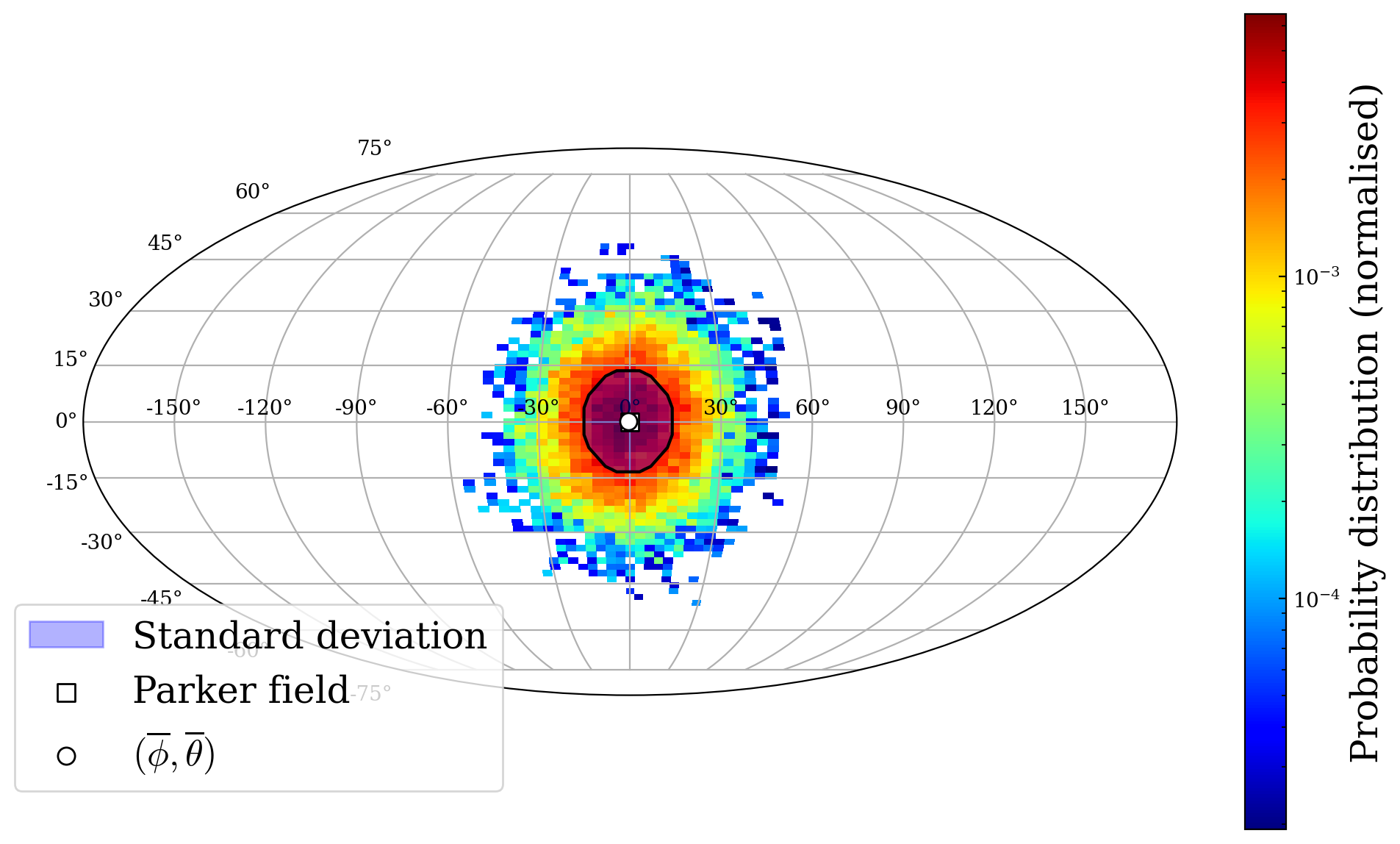}
\includegraphics[width=0.5\textwidth]{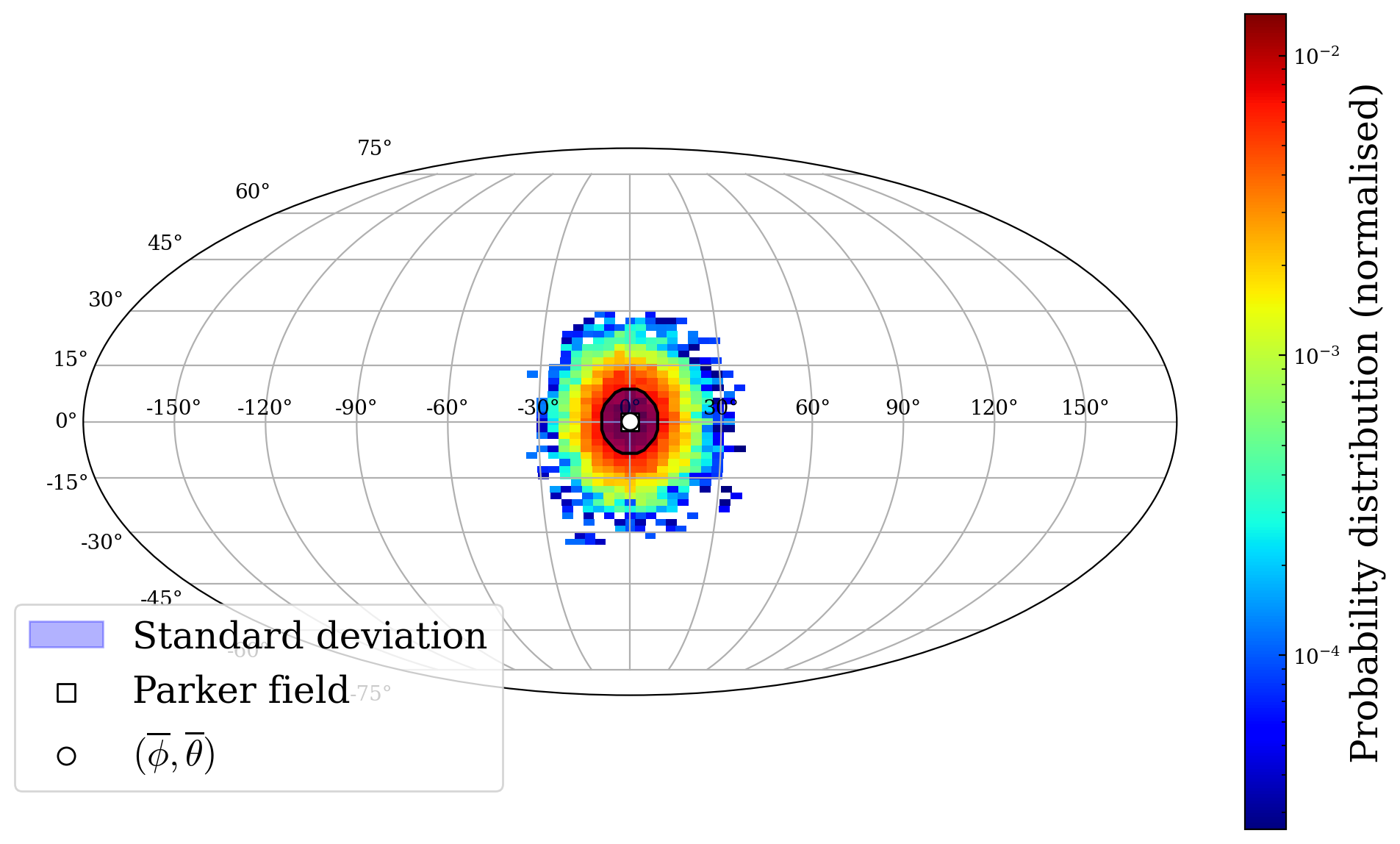}
\includegraphics[width=0.5\textwidth]{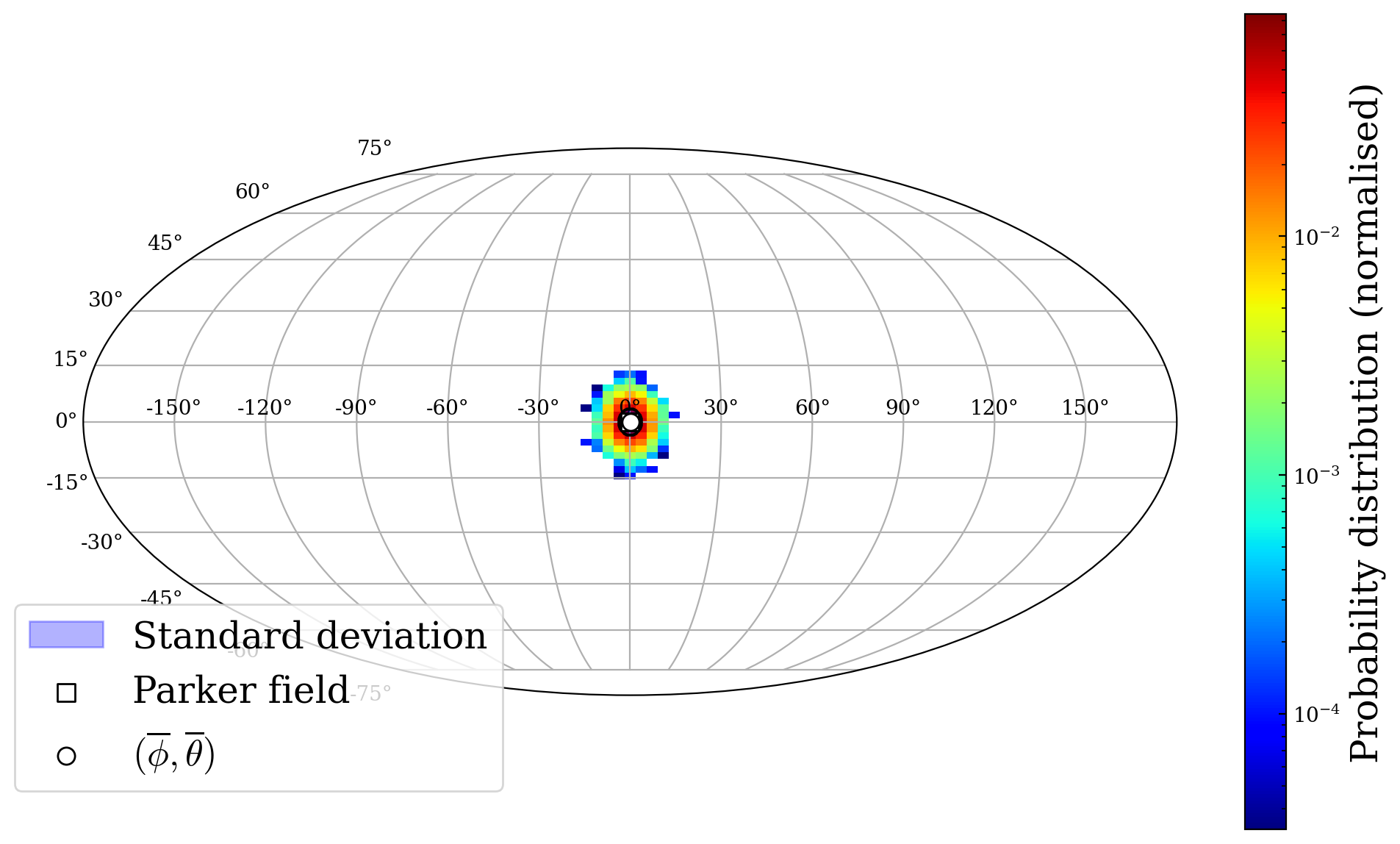}
\caption{Mollweide projections depicting the shifted probability distributions at a radial distance of $0.04$ AU from the Sun obtained using the time-backwards formulations with starting radii 1 AU (top left), 0.7 AU (top right), 0.5 AU (bottom left) and 0.25 AU (bottom right). The $(\overline{\phi},\overline{\theta})$ value (white circle), Parker spiral (white square) along with the $\sigma$ value (blue area) are indicated once more on each of the projections. \label{fig:backwards mollweide_2}}
\end{figure*}

Similar to Fig.~\ref{fig:mollweide}, Fig.~\ref{fig:backwards mollweide_2} shows four Mollweide projections of the probability distribution at the inner boundary, located $0.04$ AU from the Sun. The projections shown are obtained by utilising the time-backward formulation, the only difference being the starting radius from where the distribution is traced back from. The respective starting radii {here} being 1 AU, 0.7 AU, 0.5 AU and 0.25 AU. Moreover, each projection in Fig.~\ref{fig:backwards mollweide_2} also displays the locations of the most likely path with respect to the Parker spiral as well as the standard deviation on the calculated most likely path. For magnetic fieldlines observed at fixed points throughout the heliosphere, Fig.~\ref{fig:backwards mollweide_2} shows the probability of where they originated from at the inner boundary. The implications thereof are discussed more in the next section.\\



\section{Discussion}\label{Sec:discussion}

In this work we simulated the stochastic meandering of Parker magnetic fieldlines to study the diffusion of fieldlines in the turbulent solar wind. Not only is the transport of energetic particles affected by these turbulent magnetic fields, but the magnetic connectivity between two points can also be significantly altered. This can have significant implications for studies assuming turbulence-free magnetic fields to extrapolate the level of magnetic connectivity, e.g. \cite{posner_hohmannparker_2013} and \citet{SteynEA24}. \\


From Fig.~\ref{fig:hist} it can be inferred that the probability distribution is more narrow close to the Sun, and spreads out as the radial distance increases. Additionally, the probability distribution spreads out faster at smaller radial distances, but the rate of increase slows down as the radial distance increases. As expected, the peak of the distribution follows the curvature of the Parker HMF, changing in $\phi$, but with $\theta$ remaining centered on the equatorial plane.\\

By calculating $\overline{\phi}(r)$ and $\overline{\theta}(r)$ at regular radial distance intervals, it becomes possible to estimate the most likely path which a magnetic fieldline is expected to follow, as illustrated in Fig.~\ref{fig:stand dev}. Both the $\overline{\phi}(r)$ and the $\overline{\theta}(r)$ values are nearly identical to the Parker spiral values. The  $\overline{\phi}(r)$ values decrease at a constant rate while remaining near the Parker spiral $\phi$ values, though the  $\overline{\phi}(r)$ values show a slower decline than those of the Parker spiral $\phi$ values. The difference in the slope for the Parker spiral $\phi$ values and the $\overline{\phi}(r)$ could imply that the most likely path obtained through the simulation would be underwound relative to what is expected of the Parker spiral, as also discussed in the work of \citet{bian_heliospheric_2024}, since most of the  $\overline{\phi}(r)$ values possess a value either equal to or greater than the Parker spiral $\phi$ values.\\

From Figs.~\ref{fig:stand dev} and \ref{fig:2D B-field grid} it can be reasoned that a field-line will most likely be found near the Parker spiral, especially at small radial distances, seeing as the $(\overline{\phi},\overline{\theta})$ values in Fig.~\ref{fig:2D B-field grid} remain near the Parker spiral, with most being on top of the spiral. It should be emphasized that, while the estimated path derived from the peaks of the probability distribution is the most likely path, it is still possible that the fieldline can take an entirely different path than the estimated one, albeit with a lower probability. It is apparent from Fig.~\ref{fig:mollweide} that the Parker spiral and the most likely path are both located very close to one another. Furthermore, these values are all located within one standard deviation, which is about $25^{\circ}$. Fig.~\ref{fig:mollweide} therefore gives an indication of the level of spreading of fieldlines diffusing from a point-source near the Sun to Earth. {Our results are very similar to the those presented by \citet{Chhiber_etal_2021}, reporting a root-mean-squared spreading, at 1 AU, of $20^{\circ} - 60^{\circ}$, implying that fieldlines, originating within the same source region, can develop
separations as wide as $120^{\circ}$ in longitude.} {Following a different approach, \cite{laitinen_analytical_2023} found that the standard deviation on the angular spread of fieldlines at 1 AU is $14^{\circ}$, a difference of about a factor 2 from our standard deviation. However, the standard deviation, as well as the spread of fieldlines, is highly sensitive with regards to the input parameters, as seen in Fig.~\ref{fig:multi}, and these are not identical in the different studies.} While the probability of finding a fieldline at the expected position of the Parker HMF is highest, the distribution spans approximately $120^{\circ}$ in either direction, and roughly 30\% of fieldlines will be more than one standard deviation, i.e. more than $25^{\circ}$, away from the average values. This has significant implications for the transport of solar energetic particles (SEPs) accelerated close to the Sun and following these random walking magnetic fieldlines into the heliosphere: just by following the diffusing fieldlines, SEPs originating from a point-source near the Sun, can spread $120^{\circ}$ in longitude. If pitch-angle scattering can lead to the decoupling of SEPs from fieldlines \cite[see, e.g.,][]{Engelbrecht19}, the actually SEP distribution could be even wider \citep[e.g.][]{strauss_perpendicular_2018,Chhiber_etal_2021b}. \\

From a cursory glance of Fig.~\ref{fig:multi mean} it would appear that all of the simulations will produce an underwound magnetic field-line, assuming the magnetic field-line perfectly follows the most likely path, since each of the $\overline{\phi}(r)$ values are located above that of the Parker spiral, with the magnetic field-line produced by simulation 4 being the most underwound. This confirms the earlier simulations of \citet{bian_heliospheric_2024}. The most noticeable difference between each graph in Fig.~\ref{fig:multi} is the increase in area of where the solution curves, and hence the HMF, can be found. For simulations 1 and 2, which have lower diffusion coefficient values, the solution curves remain closer to the Parker spiral for greater distances. On the other hand, simulations 3 and 4 produce a much larger area as a result of the larger diffusion coefficient values. For large enough diffusion coefficient values the solution curves, which usually show the same tendency as the Parker spiral, will be dispersed throughout the entire area around the Sun as seen in the right-most graph of Fig.~\ref{fig:multi}. Such large values for the fieldline diffusion coefficient may be somewhat extreme, but not unrealistic: In this work we used turbulence parameters derived mostly from quiet-time solar minimum plasma quantities which could, in principle, be significantly more disturbed during solar maximum or due to other, transient, conditions.\\

Upon closer examination of Fig.~\ref{fig:multi}, it becomes evident that it is not just the solution curve trajectories and the area in which the HMF can be found which are changed, but the most likely path as well. The most likely path yielded by simulation 1 possesses close to the same curvature as the Parker spiral, although it shows a hint of being overwound, albeit within the error margins. Simulation 2, on the other hand, produces a most likely path almost identical to the Parker spiral. In contrast, simulations 3 and 4 show most likely paths which are clearly underwound, with simulation 4 showing the greatest change in the most likely path. Whereas simulations 1, 2 and 3 yield most likely paths which seem to show a similar curvature to that of the Parker spiral, simulation 4 only starts out with a similar curvature; after about $0.9$ AU the most likely path seems to almost straighten out and become almost radial. Consequently, the most likely path, and hence the HMF, will be similar to the Parker spiral for less diffusion, while underwound for greater diffusion. This corresponds with the observational analysis of winding angles at $1$~AU of \cite{vander_merwe_revisiting_2025}. As discussed in \citet{bian_heliospheric_2024}, this underwinding is due to a combination of the curvature of the background Parker HMF and the spatial dependence of the fieldline diffusion coefficient. The underwinding is thus a natural consequence of the presence of turbulence and increases with the assumed level of turbulence. \\

The first result obtained from the backward formulation shows a strong resemblance to the results obtained from the forward formulation, as is evident when comparing Fig.~\ref{fig:mollweide} with the top left graph of Fig.~\ref{fig:backwards mollweide_2}. Therefore, it can be argued that any magnetic fieldline observed at the Earth and traced back towards the Sun most likely possesses an origin somewhere within one standard deviation of what one would expect of the Parker spiral. Although with a lesser probability, it is still possible that the fieldline originates outside of the standard deviation and meanders towards the Earth due to the FLRW process. Fig.~\ref{fig:backwards mollweide_2} therefore suggest that, just due to the presence of turbulence and fieldline diffusion, ballistic back-mapping of magnetic fieldlines from Earth back towards the Sun introduces an uncertainty of around $25^{\circ}$ (roughly one standard deviation of the probability distribution) while, on occasion, the deviation could potentially be much larger. {It is important to note that the spreading of fieldlines due to the FLRW process is an intrinsic source of uncertainty originating from the presence of solar wind turbulence. There are also model uncertainties resulting from the back-mapping of interplanetary magnetic fieldlines to the Sun \citep[e.g.][]{daSilva_etal_2023}.} In addition, Fig.~\ref{fig:backwards mollweide_2} shows that if the fieldline is observed at smaller radial distances this uncertainty decreases, meaning that fieldlines observed by spacecraft close to the Sun, such as the Parker Solar Probe or Solar Orbiter, can be traced back to their origin with more certainty.\\


\appendix
\section{Numerical aspects} \label{Sec:num_aspects}
{A convenient numerical method which can be used to solve partial differential equations involving diffusive processes is the stochastic differential equation (SDE) formulation \citep[][]{strauss_hitch-hikers_2017}. In the one-dimensional case an SDE takes the general form}
\begin{equation}
    \frac{dx(t)}{dt} = a(x,t) + b(x,t)\xi(t), \label{eq:SDE definition}
\end{equation}
{where the functions $a(x,t)$ and $b(x,t)$ are both continuous and the white noise term, $\xi(t)$, is a rapidly varying stochastic term. The drift term, $a(x,t)$, accounts for the deterministic component and the diffusion term,  $b(x,t)\xi(t)$, for the stochastic component of the system, therefore also including the diffusive processes. There is a well established mathematical equivalence between any second-order partial differential equation and it's equivalent stochastic formulation \citep[][]{gardiner_handbook_1985, kloeden_numerical_1999, lemons_introduction_2002, van_kampen_stochastic_2007}.} 

The heliospheric magnetic field lines $x_{1}(s)=r(s)$, $x_{2}(s)=\theta(s)$ and $x_{3}(s)=\phi(s)$ are modeled by a continuous Markov process closely related to the one underlying the focused transport equation. The difference is that we only consider a \emph{subclass} of sample realizations, those with $\mu=1$: the scatter-free process. It can be looked at in a ``forward'' or ``backward'' way depending whether its transition probability density is the solution of the Kolmogorov forward (Fokker-Planck) or Kolmogorov backward equations \citep{gardiner_handbook_1985, kloeden_numerical_1999, oksendal_stochastic_2000, lemons_introduction_2002, van_kampen_stochastic_2007}. We deal with SDE's of the form ,
\begin{equation}\label{sde0}
dx_i(s)=a_i(x_i,s)ds + \sum_{j=1}^{3}b_{ij}(x_i,s)dW_i(s),
\end{equation}
which are interpreted in the the sense of Ito. Eq. (\ref{sde0}) can be solved iteratively by using its discrete, Euler-Maruayama version, given by
\begin{equation}\label{eq:iterate}
	x_{i}(s+\Delta s) = x_{i}(s)+a_{i}(x_{i},s)\Delta s+\sum_{j=1}^{3}b_{ij}(x_i,s)\xi_{i},
\end{equation}
where $\xi_{i}=\Delta W_{i}$ are mutually independent centered Gaussian random variables with variance $\Delta s$.

\subsection{Forward approach}

In the forward version of the model, the convection-diffusion equation
is transformed into a Fokker-Planck equation for the probability density $g_m=f_mr^2\sin\theta$, yielding 
\begin{eqnarray}
    \frac{\partial g_m}{\partial s} &=& \frac{\partial^2}{\partial r^2}\left(D_{\perp}\sin^2\psi g_m\right)+\frac{\partial^2}{\partial\theta^2}\left(\frac{D_{\perp}}{r^2}g_m\right) +\frac{\partial^2}{\partial\phi^2}\left(\frac{D_{\perp}\cos^2\psi}{r^2\sin^2\theta}g_m\right)+2\frac{\partial^2}{\partial r\partial\phi}\left(\frac{D_{\perp}\cos\psi\sin\psi}{r\sin\theta}g_m\right) \nonumber\\
    &-& \frac{\partial}{\partial r}\left(\left[\frac{\partial}{\partial r}(D_{\perp}\sin^2\psi)+\frac{2}{r}D_{\perp}\sin^2\psi+\cos\psi\right]g_m\right)-\frac{\partial}{\partial\theta}\left(\left[\frac{\partial}{\partial\theta}\left(\frac{D_{\perp}}{r^2}\right)+\frac{D_{\perp}}{r^2}\cot\theta\right]g_m\right)\nonumber\\
    &-&\frac{\partial}{\partial\phi}\left(\left[\frac{D_{\perp}\cos\psi\sin\psi}{r^2\sin\theta}\right.\right.\left.\left.+\frac{1}{r\sin\theta}\frac{\partial}{\partial r}(D_{\perp}\cos\psi\sin\psi)-\frac{\Omega}{v_{sw}}\cos\psi\right]g_m\right),\label{eq:super conservative}
\end{eqnarray}
in a spherical coordinate system. This equation can be written in the general form
\begin{equation}
	\frac{\partial g_m}{\partial s} = -\sum_{i=1}^{n}\frac{\partial}{\partial x_i}\left[a_ig_m\right]+\frac{1}{2}\sum_{i=1}^{n}\sum_{j=1}^{n}\frac{\partial^2}{\partial x_i\partial x_j}\left[C_{ij}g_m\right]+\mathcal{L}g_m+\mathcal{Q},\label{eq:forward kol}
\end{equation}
 where $a_i$ represents the deterministic drift terms and $b_{ij}$ is a positive definite matrix related to the diffusion tensor $C_{ij}$ by $b_{ij} \cdot b_{ij}^T = C_{ij}$. {This general formulation allows for the inclusion of possible linear $(\mathcal{L})$ or source $(\mathcal{Q})$ terms, both of which are zero for the forward approach discussed here, but do arise in, for example, the backward formulation discussed in the next paragraph.} Therefore, the set of SDEs to be solved are given by 
\begin{eqnarray}
    dx_i (s + ds) &=& a_i (s) ds + \sum_{j=1}^{n}b_{ij} (s) dW_i (s)   \nonumber \\
    a(s + ds) &=& a (s) + Q(s) ds \nonumber \\
    w(s + ds) &=& w_0\exp \left\{ \mathcal{L}(s) ds \right\},
\end{eqnarray}
where
\begin{eqnarray}
    a_r &=&\left[\frac{\partial}{\partial r}\left(D_{\perp}\sin^2\psi\right)+\frac{2}{r}D_{\perp}\sin^2\psi+\cos\psi\right],\nonumber\\\nonumber\\
    a_{\theta} &=& \left[\frac{1}{r^2}\frac{\partial}{\partial\theta}\left(D_{\perp}\right)+\frac{D_{\perp}}{r^2}\cot\theta\right],\nonumber\\\nonumber\\
    a_{\phi}&=&\left[\frac{D_{\perp}\cos\psi\sin\psi}{r^2\sin\theta}+\frac{1}{r\sin\theta}\frac{\partial}{\partial r}\left(D_{\perp}\cos\psi\sin\psi\right)-\frac{\Omega}{v_{sw}}\cos\psi\right],\nonumber
\end{eqnarray}
$$\textbf{b} = \sqrt{2}\left[\begin{array}{ccc}
     \sqrt{\frac{D_{rr}D_{\theta\phi}^2-D_{rr}D_{\theta\theta}D_{\phi\phi}-2D_{r\phi}D_{\theta\phi}D_{r\theta}+D_{r\theta}^2D_{\phi\phi}+D_{r\phi}^2D_{\theta\theta}}{D_{\theta\phi}^2-D_{\theta\theta}D_{\phi\phi}}} & \frac{D_{r\phi}D_{\theta\phi}-D_{r\theta}D_{\phi\phi}}{D_{\theta\phi}^2-D_{\theta\theta}D_{\phi\phi}}\sqrt{D_{\theta\theta}-\frac{D_{\theta\phi}^2}{D_{\phi\phi}}} & \frac{D_{r\phi}}{\sqrt{D_{\phi\phi}}}\\\\
	0&\frac{1}{r}\sqrt{D_{\theta\theta}-\frac{D_{\theta\phi}^2}{D_{\phi\phi}}}& \frac{D_{\theta\phi}}{r\sqrt{D_{\phi\phi}}}\\\\
	0&0&\frac{\sqrt{D_{\phi\phi}}}{r\sin\theta}
\end{array}\right],$$
in the general case. It reduces to 
\begin{eqnarray}
\textbf{b} = \sqrt{2 D_{\perp}(r)}\left[\begin{array}{ccc}
		0 & \qquad0 & \qquad \sin\psi\\
		0 & \qquad \frac{1}{r}& \qquad 0\\
		0 & \qquad 0 & \qquad \frac{\cos\psi}{r\sin\theta}
	\end{array}\right],\label{b matrix}
\end{eqnarray}
when $D_{||} = 0$. Additionally, $\mathcal{Q} = \mathcal{L} = 0$, as already mentioned, in the forward simulations. The set of SDEs is solved iteratively with initial condition $\left(r_0, \phi_0, \theta_0, a_0, w_0 \right) = \left(0.04 \, \mathrm{AU}, \pi, \pi/2, 0, 1 \right)$. Each simulation run by the solver consists of 10000 solutions, each consisting of 10000 steps with a step size of 0.0005 AU. The spatial trajectory $x_i$ represents the solution curve from a single realization of the set of equations. A single realization is statistically meaningless, so a large number of realizations are simulated. The numerical solution can only be obtained along these curves. If no curves pass through a point, the solution at that point is undefined. Along the curve, each realization contributes an amount $a + w$ to the total solution, which is the probability distribution calculated by averaging the contributions from all SDE realizations passing through a region of interest. {For this forward simulation where $\mathcal{Q} = \mathcal{L} = 0$ we have $a = 0$ and $w=1$, so that each particle contributes an equal weight of $a + w = 0 + 1$ to the average. When either a linear or source term is present, as will be the case in the next paragraph, these weights will deviate from unity. }
\subsection{Backwards simulations}

In contrast to the forward simulations, in the backwards simulations one begins with an initial condition inside the heliosphere, say at 1 AU, and solves the relevant set of SDEs backwards until the inner boundary close to the Sun is reached. In this formulation Eq.~\ref{eq:convection diff} needs to be cast into a non-conservative form

\begin{eqnarray}
    \frac{\partial f_m}{\partial s^{\prime}} &=& \left[\frac{2}{r}D_{\perp}\left(\sin^2\psi+\tan^2\psi\left[1+\tan^2\psi\right]^{-2}\right)\right.\left.+\sin^2\psi\frac{\partial D_{\perp}}{\partial r}-\cos\psi\right]\frac{\partial f_m}{\partial r} \\
    &+&\left[\frac{\cos\psi\sin\psi}{r^2\sin\theta}\left(D_{\perp}+r\frac{\partial D_{\perp}}{\partial r}+D_{\perp}\left[\cos^2\psi-\sin^2\psi\right]\right) +\frac{\Omega}{v_{sw}\cos\psi}\right]\frac{\partial f_m}{\partial\phi} \nonumber \\
    &+&D_{\perp}\sin^2\psi\frac{\partial^2f_m}{\partial r^2}+\frac{D_{\perp}}{r^2}\frac{\partial^2f_m}{\partial\theta^2}+\frac{D_{\perp}\cos^2\psi}{r^2\sin^2\theta}\frac{\partial^2f_m}{\partial\phi^2}+\frac{2D_{\perp}\cos\psi\sin\psi}{r\sin\theta}\frac{\partial^2f_m}{\partial r\partial\phi} \nonumber \\
    &+& \left[\frac{1}{r^2}\left(\frac{\partial D_{\perp}}{\partial\theta}+D_{\perp}\cot\theta\right)\right]\frac{\partial f_m}{\partial\theta} - \left(\frac{2+\tan^2\psi}{r}\cos^3\psi\right)f_m\label{eq:backwards formulation} \nonumber
\end{eqnarray}
which possesses the general form
\begin{equation}
    -\frac{\partial f_m}{\partial s^{\prime}} = \widetilde{a}_i\frac{\partial f_m}{\partial x_i} + \frac{1}{2}\widetilde{C}_{ij}\frac{\partial^2f_m}{\partial x_i^2} + \mathcal{L} f_m + \mathcal{Q}, \label{Eq:general_cons_backwards}
\end{equation}
thus the set of SDEs for the time-backward formulation is given by
\begin{equation}
	-\frac{\partial f_m}{\partial s^{\prime}}=\sum_{i=1}^{n}\widetilde{a}_i\frac{\partial f_m}{\partial x_i}+\frac{1}{2}\sum_{i=1}^{n}\sum_{j=1}^{n}\widetilde{C}_{ij}(x_i,s)\frac{\partial^2 f_m}{\partial x_i\partial x_j}+\mathcal{L}(x_i,s)f_m + \mathcal{Q}\label{eq:backward kol}
\end{equation}
which is the time backwards Kolmogorov equation. For the time-backward formulation, the set of SDEs is

\begin{eqnarray}
    a_r &=&\frac{2}{r}D_{\perp}\left(\sin^2\psi+\tan^2\psi\left[1+\tan^2\psi\right]^{-2}\right)+\sin^2\psi\frac{\partial D_{\perp}}{\partial r}-\cos\psi\nonumber\\\nonumber\\
    a_{\theta} &=&\frac{1}{r^2}\left(\frac{\partial D_{\perp}}{\partial\theta}+D_{\perp}\cot\theta\right) \nonumber\\\nonumber\\
    a_{\phi}&=&\frac{\cos\psi\sin\psi}{r^2\sin\theta}\left(D_{\perp}+r\frac{\partial D_{\perp}}{\partial r}+D_{\perp}\left[\cos^2\psi-\sin^2\psi\right]\right)+\frac{\Omega}{v_{sw}}\cos\psi\nonumber\\\nonumber\\
    \mathcal{L} &=& -\frac{2+\tan^2\psi}{r}\cos^3\psi \nonumber\\
    \mathcal{Q} &=& 0,
\end{eqnarray}
where $\vec{b}$ once again reduces to Eq.~\ref{b matrix}. For the time-backward formulation the set of SDEs is solved iteratively from an initial condition $\left(r_0, \phi_0, \theta_0, a_0, w_0 \right) = \left(1 \, \mathrm{AU}, 2.12, \pi/2, 0, 1 \right)$. {As there is a linear term present, the particle weight, $w$, changes with distance and different SDE realization contribute differently to the averages.} Each simulation run by the solver consists of 10000 solutions, each consisting of a possible 10000 steps with a step size of 0.0005 AU. While each solution can consist of 10000 step, the solver stops the calculation once $r$ reaches a value of 0.04 AU.\\

\subsection{The diffusion tensor in spherical coordinates}

The background HMF is the Parker field. In terms of a local field aligned coordinate system,
$$\textbf{D}=\left[\begin{array}{c c c}
     D_{||} & 0 & 0  \\
     0 & D_{\perp} & 0 \\
     0 & 0 & D_{\perp }
\end{array}\right],$$
where $\hat{e}_{\parallel}$ is parallel to the HMF, $\hat{e}_1$ is perpendicular to $\hat{e}_{\parallel}$ and $\hat{e}_2$ lies in  $r\phi$. In the global spherical coordinate system, the local field aligned coordinate system can be expressed as
\begin{eqnarray}
    \hat{e}_{||} &=& \cos\psi\hat{e}_r - \sin\psi\hat{e}_{\phi},\nonumber\\
	\hat{e}_1 &=& \hat{e}_{\theta},\nonumber\\
	\hat{e}_2 &=& \hat{e}_{||} \times \hat{e}_1 = \sin\psi\hat{e}_r + \cos\psi\hat{e}_{\phi}, 
\end{eqnarray}
where $\psi$ is the HMF winding angle. Therefore, the transformation matrix from the local to the global coordinate system is given by
$$\textbf{T}=\left[\begin{array}{c c c}
     \cos\psi & 0 & \sin\psi \\
	0 & 1 & 0 \\
	-\sin\psi & 0 & \cos\psi \\
\end{array}\right],$$
where det$(\textbf{T}) = 1$. Therefore, $\textbf{D}$ can be written in spherical coordinates as

\begin{eqnarray}
    \left[\begin{array}{ccc}
        D_{rr} & D_{r\theta} & D_{r\phi}\\
		D_{\theta r} & D_{\theta\theta} & D_{\theta\phi}\\
		D_{\phi r} & D_{\phi\theta} & D_{\phi\phi} 
    \end{array}\right] = \textbf{T}\,\textbf{D}\,\textbf{T}^T
    &=&\left[\begin{array}{ccc}
        \cos\psi & 0 & \sin\psi\\
		0 & 1 & 0\\
		-\sin\psi & 0 & \cos\psi
    \end{array}\right]\left[\begin{array}{ccc}
        D_{||} & 0 & 0\\
		0 & D_{\perp} & 0\\
		0 & 0 & D_{\perp r} 
    \end{array}\right]\left[\begin{array}{ccc}
        \cos\psi & 0 & -\sin\psi\\
		0 & 1 & 0\\
		\sin\psi & 0 & \cos\psi
    \end{array}\right]\nonumber\\
    &=&\left[\begin{array}{ccc}
        D_{||}\cos^2\psi + D_{\perp }\sin^2\psi & 0 & (D_{\perp }-D_{||})\cos\psi\sin\psi\\
		0 & D_{\perp} & 0\\
		(D_{\perp }-D_{||})\cos\psi\sin\psi & 0 & D_{||}\sin^2\psi + D_{\perp }\cos^2\psi
    \end{array}\right]\label{eq:D matix}
\end{eqnarray}

where $\textbf{T}^T$ is the transpose of the transformation matrix.


\bibliography{Masters}{}
\bibliographystyle{aasjournalv7}



\end{document}